\documentclass[final,3p,times,twocolumn]{elsarticle}
\usepackage{amssymb}
\usepackage[usenames,dvipsnames]{xcolor}

\usepackage{chemscheme}

\usepackage{chemformula} % Formula subscripts using \ch{}
\biboptions{sort&compress}

\journal{Journal of Colloid and Interface Science}

\begin{document}

\begin{frontmatter}

%% Title, authors and addresses

%% use the tnoteref command within \title for footnotes;
%% use the tnotetext command for theassociated footnote;
%% use the fnref command within \author or \address for footnotes;
%% use the fntext command for theassociated footnote;
%% use the corref command within \author for corresponding author footnotes;
%% use the cortext command for theassociated footnote;
%% use the ead command for the email address,
%% and the form \ead[url] for the home page:
%% \title{Title\tnoteref{label1}}
%% \tnotetext[label1]{}
%% \author{Name\corref{cor1}\fnref{label2}}
%% \ead{email address}
%% \ead[url]{home page}
%% \fntext[label2]{}
%% \cortext[cor1]{}
%% \affiliation{organization={},
%%             addressline={},
%%             city={},
%%             postcode={},
%%             state={},
%%             country={}}
%% \fntext[label3]{}

\title{Simulation of the Carbon Dioxide Hydrate-Water Interfacial
Energy}

%% use optional labels to link authors explicitly to addresses:
%% \author[label1,label2]{}
%% \affiliation[label1]{organization={},
%%             addressline={},
%%             city={},
%%             postcode={},
%%             state={},
%%             country={}}
%%
%% \affiliation[label2]{organization={},
%%             addressline={},
%%             city={},
%%             postcode={},
%%             state={},
%%             country={}}

\author[inst1]{Jes\'us Algaba}
\author[inst2]{Esteban Acu\~na}
\author[inst2]{Jos\'e Manuel M\'iguez}
\author[inst3]{Bruno Mendiboure}
\author[inst2]{Iv\'an M. Zer\'on}
\author[inst2]{Felipe J. Blas\corref{mycorrespondingauthor}}
\cortext[mycorrespondingauthor]{Corresponding author}
\ead{felipe@uhu.es}

\affiliation[inst1]{organization={Department of Chemical Engineering, South Kensington Campus, Imperial College London, SW7 2AZ London (United Kingdom)},%Department and Organization
%            addressline={Address One}, 
%            city={City One},
%            postcode={00000}, 
%            state={State One},
%            country={Country One}}
}

\affiliation[inst2]{organization={Laboratorio de Simulacion Molecular y Quimica Computacional, CIQSO-Centro de Investigacion en Quimica Sostenible and Departamento de Ciencias Integradas, Universidad de Huelva, 21007-Huelva (Spain)},%Department and Organization
%            addressline={Address Two}, 
%            city={City Two},
%            postcode={22222}, 
%            state={State Two},
%            country={Country Two}}
}

\affiliation[inst3]{organization={Laboratoire des Fluides Complexes et Leurs Reservoirs, UMR5150, Universite de Pau et des Pays de l'Adour, B. P. 1155, Pau Cdex 64014 (France)}%Department and Organization
%            addressline={Address Two}, 
%            city={City Two},
%            postcode={22222}, 
%            state={State Two},
%            country={Country Two}}
}

\begin{abstract}
%% Text of abstract

\noindent
\emph{Hypothesis:} Carbon dioxide hydrates are ice-like nonstoichiometric inclusion solid compounds with importance to global climate change, and gas transportation and storage. The thermodynamic and kinetic mechanisms that control carbon dioxide nucleation critically depend on hydrate-water interfacial free energy. Only two independent indirect experiments are available in the literature. Interfacial energies show large uncertainties due to the conditions at which experiments are performed. Under these circumstances, we hypothesize that accurate molecular models for water and carbon dioxide combined with computer simulation tools can offer an alternative but complementary way to estimate interfacial energies at coexistence conditions from a molecular perspective.

\noindent
\emph{Calculations:} We have evaluated the interfacial free energy of carbon dioxide hydrates at coexistence conditions (three-phase equilibrium or dissociation line) implementing advanced computational methodologies, including the novel Mold Integration methodology. Our calculations are based on the definition of the interfacial free energy, standard statistical thermodynamic techniques, and the use of the most reliable and used molecular models for water (TIP4P/Ice) and carbon dioxide (TraPPE) available in the literature.

\noindent
\emph{Findings:} We find that simulations provide an interfacial energy value, at coexistence conditions, consistent with the experiments from its thermodynamic definition. Our calculations are reliable since are based on the use of two molecular models that accurately predict: (1) The ice-water interfacial free energy; and (2) the dissociation line of carbon dioxide hydrates. Computer simulation predictions provide alternative but reliable estimates of the carbon dioxide interfacial energy. Our pioneering work demonstrates that is possible to predict interfacial energies of hydrates from a truly computational molecular perspective and opens a new door to the determination of free energies of hydrates.   

\end{abstract}

%%Graphical abstract
\begin{graphicalabstract}
\includegraphics[angle=0,width=\textwidth]{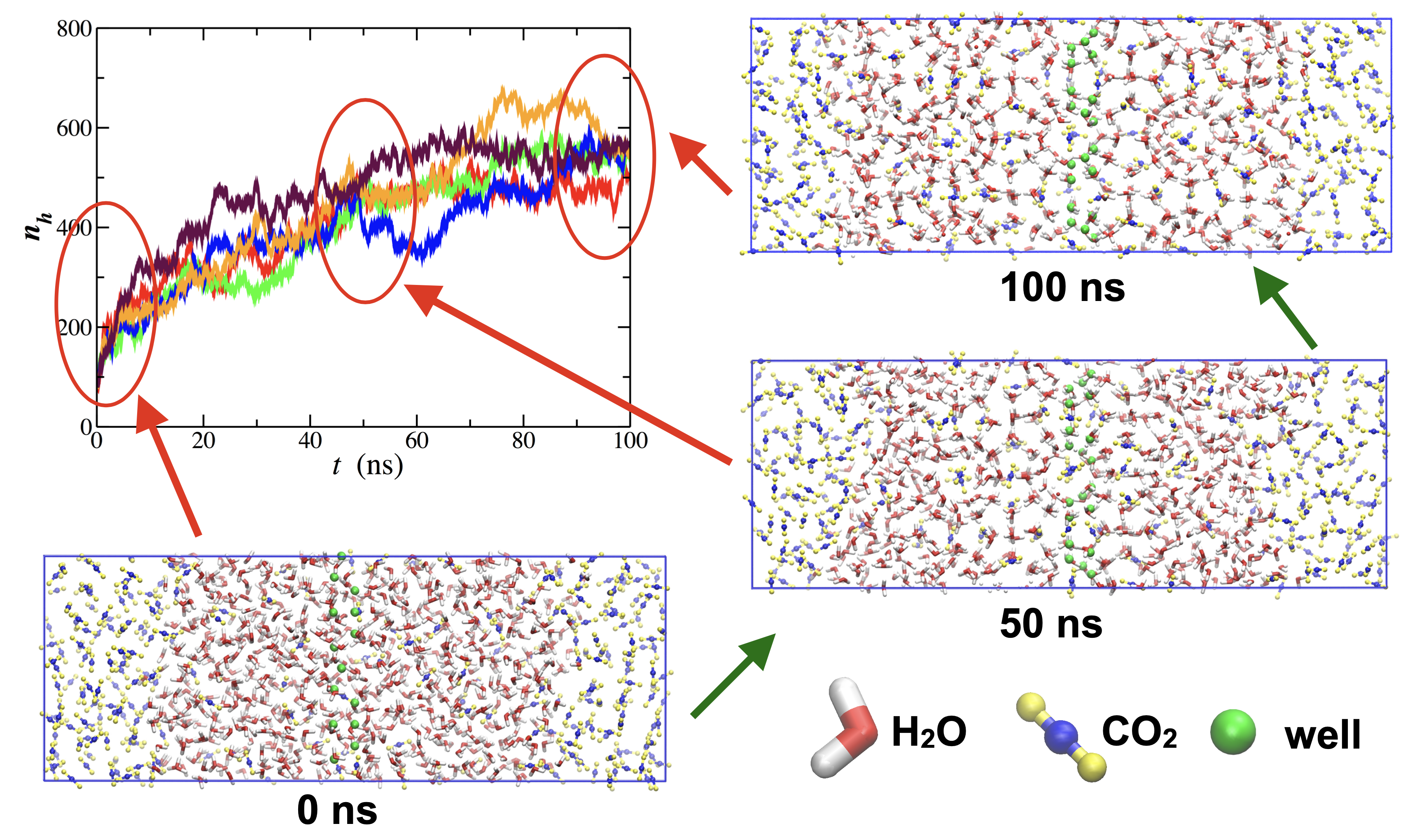}
\end{graphicalabstract}

%%Research highlights
%\begin{highlights}
%\item Research highlight 1
%\item Research highlight 2
%\end{highlights}

\begin{keyword}
%% keywords here, in the form: keyword \sep keyword

Interfacial free energy \sep carbon dioxide hydrate \sep
hydrate-water interface \sep computer simulation

%% PACS codes here, in the form: \PACS code \sep code
%%\PACS 0000 \sep 1111
%% MSC codes here, in the form: \MSC code \sep code
%% or \MSC[2008] code \sep code (2000 is the default)
%%\MSC 0000 \sep 1111
\end{keyword}

\end{frontmatter}

%% \linenumbers

%% main text

%%%%%%%%%%%%%%%%%%%%%%%%%%%%%%%%%%%%%%%%%%%%%%%%%%%%%%%%%%%%%%%%%%%%%
%% Start the main part of the manuscript here.
%%%%%%%%%%%%%%%%%%%%%%%%%%%%%%%%%%%%%%%%%%%%%%%%%%%%%%%%%%%%%%%%%%%%%
\section{Introduction}

Hydrates are nonstoichiometric inclusion solid compounds in
which guest molecules, such as methane (\ch{CH4}), carbon dioxide
(\ch{CO2}) and molecular hydrogen (\ch{H2}), among many others species, are enclathrated in the voids left by a periodic network of water molecules (host)~\cite{Sloan2008a}. Hydrates of small molecules, such as \ch{CH4}, \ch{CO2} and \ch{H2} form sI hydrates~\cite{Sloan2008a}, while larger substances usually
form sII and sH hydrate structures~\cite{Manakov2003a,Makino2005a}. Fundamental and applied research on hydrates has been motivated by several reasons. From an applied point of view, hydrates are potential alternative sources of energy as \ch{CH4} hydrate deposits~\cite{Kvenvolden1988a,Koh2012a}, and also are important from their global climate impact~\cite{Sloan2008a,Manakov2003a,Makino2005a,Kvenvolden1988a,Koh2012a,Sloan2003a}, \ch{CO2} sequestration~\cite{Ohgaki1996a,Yang2014a,Ricaurte2014a} and gas storage~\cite{Kvamme2007a} and transportation~\cite{Chihaia2005a,Peters2008a,English2009a}.

From a more general and fundamental point of view, the
mechanisms that control the thermodynamics and growth kinetics
of these materials are still poorly understood~\cite{Sloan2008a,Debenedetti1996a}. The precise
knowledge of these parameters is essential to understand, from a
molecular perspective, the growth patterns and nucleation, and
particularly interfacial free energies of these strategic materials~\cite{Aman2016a}. The measurement or estimation of the hydrate-water interfacial energy remains a primary knowledge gap, since this property controls, not only hydrate nucleation, but also other important issues including growth mechanics and adsorption and packing of surfactants at the hydrate crystal surface~\cite{Aman2016a,Majid2021a,Hu2017a,Hu2021a}.

In this work we concentrate on the determination of interfacial free energies of type sI hydrates, and particularly on the \ch{CO2} hydrate- water interfacial free energy.

Although there exist well established and reliable methods in
the literature for the experimental determination of fluid-fluid (FF) interfacial tensions~\cite{Adamson1997a,evans200515}, the situation is not as optimistic in the case of solid-fluid (SF) interfaces. There is a rather limited number of methods for obtaining experimental free energies from experiments or theoretical calculations~\cite{Aman2016a}.

\begin{scheme}[h]
\centering
\includegraphics[angle=0,width=\columnwidth]{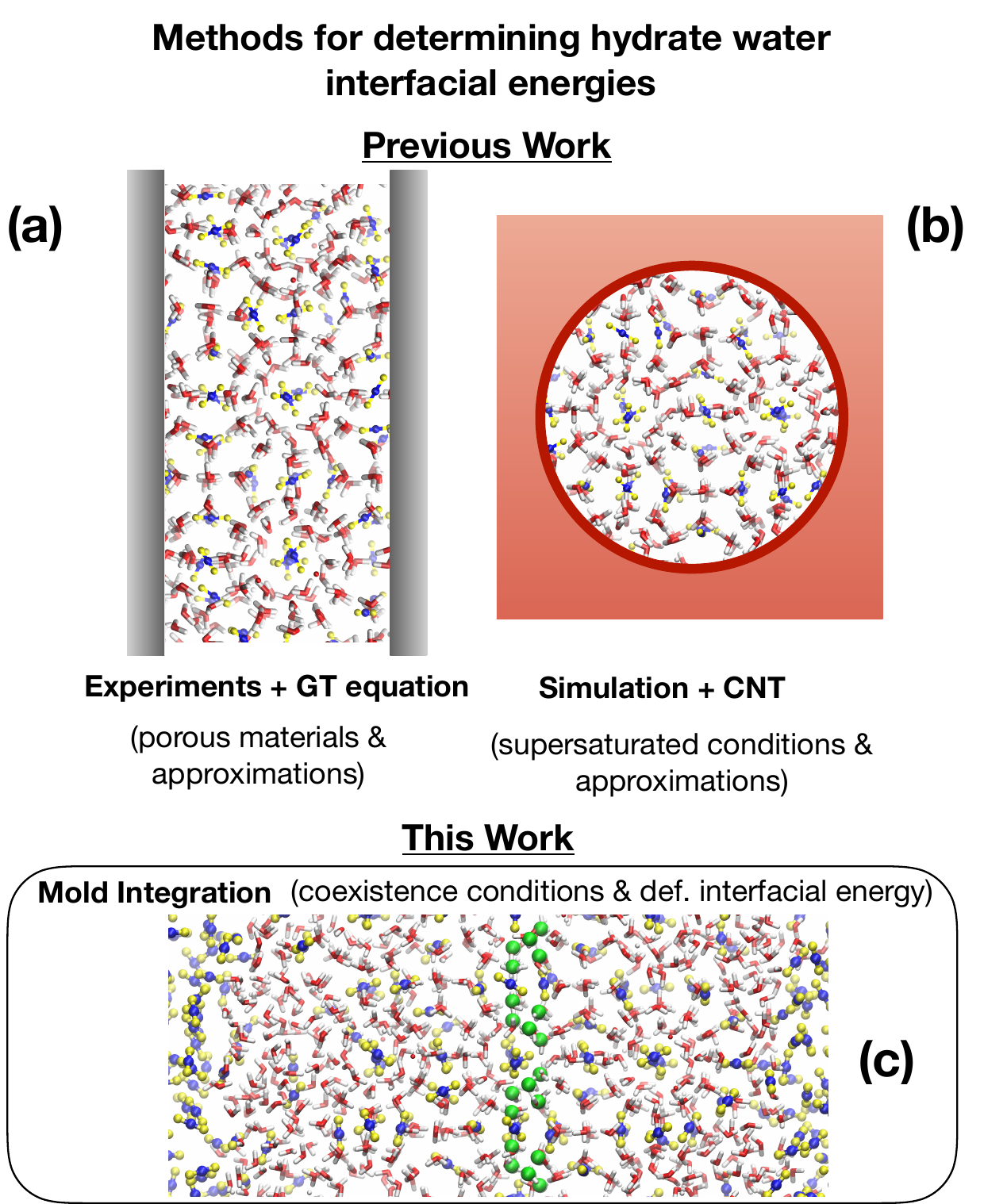}
\caption{\small Methods for determining hydrate-water interfacial energies: (a) experiments and use of the Gibbs-Thomson (GT) equation; (b) computer simulation combined with Classical Nucleation Theory (CNT); (c) Mold Integration methodology proposed in this work.}
\label{scheme}
\end{scheme}

Experimental data consists of indirect measurements
combining the difference between the melting or dissociation
temperature in pore and that in bulk of hydrates and the Gibbs-Thomson relationship~\cite{Handa1992a,Clennell1999a,Henry1999a}. The method is based on the
measurement, at constant pressure, of the dissociation/formation
of a hydrate inside porous materials with a given mean pore diameter, as indicated in part (a) of Scheme \ref{scheme}. The temperature
at which this happens depends on the pore diameter and is
different from that in bulk phase. The relative difference between
both temperatures is given in terms of different magnitudes,
including the interfacial free energy of the \ch{CO2} hydrate-water interface through the Gibbs-Thomson equation. The most important approximations of this approach are the capillary
approximation and that all pores in the porous materials are
cylindrical, with only one and well-defined pore size, independent, and nonconnected pores. Unfortunately, real mesoporous materials are far from this idyllic picture. They are made by a network of interconnected cylindrical pores with a given pore size distribution. This produces a combination of “individual” experimental dissociation curves and different definitions for the mean pore diameter, provoking large discrepancies between interfacial energy values found in the literature. To the best of our knowledge, Uchida \textit{et al.}
~\cite{Uchida1999a,Uchida2002a} and Anderson \textit{et al.}
~\cite{Anderson2003a,Anderson2003b} were the
first and only authors that measured the interfacial free energies
of \ch{CO2} hydrates using this indirect method, obtaining $28(6)$ and $30(3)~\text{mJ/m}^{2}$, respectively. Other authors have also studied the effect of different porous media on equilibrium pressure of several hydrates~\cite{Seo2002a,Zhang2003a,Smith2002a,Smith2004a,Kang2008a,Zhang2016a,Liu2018a,Zarifi2016a}.
According to this, the CO$_{2}$ hydrate-water interfacial energy ranges from $22$ to
$33\,\text{mJ/m}^{2}$. This is a large range compared, for example, with the accepted value in the literature for the surface tension of liquid water at ambient conditions, $72\,\text{mJ/m}^{2}$. The situation is similar for other solid-water interfacial energies, such as that of the ice-water interface, which ranges from $25$ to $35\,\text{mJ/m}^{2}$~\cite{Espinosa2016a}. More recently, Phan \emph{et al.}~\cite{Phan2022a} have performed a set of nice experiments to measure the cyclopentane hydrate-water interfacial energy, finding a value
of $0.24(16)\,\text{mJ/m}^{2}$.

When experimental values are scarce or it is necessary to
assess experimental measures of some magnitudes, as in the
case of the hydrate-water interfacial free energy, computer
simulations of realistic models constitute a valuable and efficient alternative method that can also improve our understanding on the hydrate-water interface from a molecular perspective~\cite{Jacobson2010a,Jacobson2010b}.
Although a number of simulation methodologies to calculate SF
interfacial free energies exists in the literature~\cite{Broughton1986a,Hoyt2001a,Angioletti-Uberti2010a,Fernandez2012a}, all of them are usually efficient for simple molecular models and not all of them are accurate, simple, and cheap from the point of view of the computational cost for complex solid structures. It is important to note that Davidchack and collaborators have determined, not only the interfacial free energy of hard-spheres~\cite{Davidchack2000a,Davidchack2010a} and Lennard-Jones~\cite{Davidchack2003a} systems using the Cleaving method, but also for several models of water~\cite{Handel2008a,Davidchack2012a}.

Several authors have determined the dissociation line of the
\ch{CO2} hydrates using computer simulation, including M\'iguez \textit{et al.}~\cite{Miguez2015a}, as well as Constandy and collaborators~\cite{Constandy2015a}, and Waage \textit{et al.}~\cite{Waage2017a}. Phan \emph{et al.}~\cite{Phan2021a} have also performed
an interesting study on the molecular mechanisms by which tetrahydrofuran affects CO$_{2}$ hydrate growth using computer simulation. However, the \ch{CO2} hydrate interfacial free energy has not been previously calculated from computer simulation at coexistence conditions. Molinero and coworkers~\cite{Jacobson2011a,Knott2012a} have
estimated indirectly the interfacial free energy of the \ch{CH4}
combining simulation and the methodologies (a) and (b) in
Scheme \ref{scheme}. However, this theory assumes that interfacial energy
of small clusters is equal to that of a macroscopic solid-liquid
planar interface. This approximation, the so-called capillarity
approximation, affects the interfacial energy values obtained from
this methodology. Other authors simulate the \ch{CH4} hydrate and
use the mechanical definition of the interfacial tension~\cite{Mirzaeifard2019a,Mirzaeifard2019b,Naeiji2017a}.
However, this methodology does not provide reliable free energy
values since the mechanical route does not account for interfacial
energies in solid phases.

Here, we propose a novel implementation of the Mold
Integration (MI) methodology proposed by Vega and
collaborators~\cite{Espinosa2016a} to deal with SF interfaces of \ch{CO2} hydrates and water, as indicated in part (c) of Scheme \ref{scheme}. The technique used in this work is based only on the thermodynamic definition of interfacial energy and on the use of standard local-order parameters. Calculation of the interfacial free energy is evaluated using a mold of attractive sites to induce reversibly the formation of a solid \ch{CO2} hydrate slab, as indicated in part (c) of Scheme \ref{scheme}.
The technique, implemented using molecular dynamics
simulations, is simpler than other methods, including the
cleaving method~\cite{Broughton1986a}, the capillary fluctuation
method~\cite{Benet2014a}, the metadynamics method~\cite{Angioletti-Uberti2010a}
or the tethered Monte Carlo method~\cite{Fernandez2012a},
and allows to predict interfacial free
energies from a fundamental point of view. The final aim of this
work is to predict the \ch{CO2} hydrate-water interfacial free energy at coexistence conditions combining this methodology with realistic and widely used water and \ch{CO2} molecular models: the
TIP4P/ice~\cite{Abascal2005b} and TraPPE~\cite{Potoff2001a}, respectively. It is important to remark two points: (1) the TIP4P/ice model is able to predict accurately the ice-water interfacial free energy~\cite{Espinosa2016a}; and (2) both models are able to quantitatively predict the three-phase or dissociation line of the \ch{CO2} hydrate~\cite{Miguez2015a}.

M\'{\i}guez and collaborators~\cite{Miguez2015a} have demonstrated that other popular
models of CO$_{2}$, including MSM~\cite{Murthy1981a,Moller1994a,Murthy1983a},
EPM2~\cite{Harris1995a}, TraPPE~\cite{Potoff2001a}, and ZD~\cite{Zhang2005a} models, in combination
with the TIP4P/Ice model for water, are able to provide
similar predictions for the dissociation line of the CO$_{2}$ hydrate (see also Figure~8
of the work of M\'{\i}guez \emph{et al.}). According to this, any of these models
for CO$_{2}$ would be appropriate to describe CO$_{2}$ hydrates. However, according
to P\'erez-S\'anchez \emph{et al.}~\cite{Perez-Sanchez2013a},
although all the models predict similar vapour-liquid equilibria
and interfacial properties of pure CO$_{2}$, the TraPPE model predicts more
accurately its complete phase diagram. In particular, this model describes
quite well the triple point temperature and pressure, as well as the melting curve. In
addition to that, the TraPPE model estimates reasonably well the melting enthalpy and
the volume change along the melting line, in better agreement with experimental
data than the other models. Due to this, M\'{\i}guez and coworkers
decided to optimized the TIP4P/Ice and TraPPE models to determine the
dissociation line of the CO$_{2}$ hydrate. We also follow the approach of these authors and
we use the TraPPE force field to model CO$_{2}$ because is the only model in the literature for which
the three-phase line of the CO$_{2}$ hydrate is available and accurately described.

It is important to mention that classical force fields for \ch{H2O} + \ch{CO2} mixture are very limited. Particularly, these models do not account for many-body contributions in polar compounds, and consequently, do not describe polarization and other electrostatic effects. During last years, several authors have described this issue~\cite{Liu2011a,Orozco2014a} and more sophisticated models including three-body interactions~\cite{Yu2012a} and the use of many body TTM-rng and MB-rng potential energy functions developed by Paesani and coworkers~\cite{Yu2012a,Riera2020a},  have been proposed. However, the use of more rigorous and detailed models requires the correct description of the \ch{CO2} hydrate dissociation line and will probably multiply the CPU time by a non-negligible factor. Particularly, the MP-pol many-body potential model for water proposed by Reddy \emph{el al.}~\cite{Reddy2016a} predicts a melting point for ice $9\,\text{K}$ lower than the experimental value. According to Conde et al.~\cite{Conde2013a}, models with a good prediction of the melting point of ice Ih (hexagonal ice) will also provide a good estimate for the dissociation line of methane hydrates, and probably the same is true for CO$_{2}$ hydrates. The main goal of this work is to show that it is possible to determine interfacial energies of hydrates from computer simulation and first principles. From this point of view, the election of a simple but enough accurate model that can be simulated using reasonably CPU resources is justified.

The novelty of our work with respect to previous results is that
we obtain the interfacial free energy of a very complex and
interesting mixture from both practical and fundamental point of
views. This is done at coexistence conditions, from fundamental
principles, including the definition of interfacial free energy and well-established tools from Thermodynamics and Statistical
Mechanics.

\section{Methods}

\textbf{Mold Integration technique.} In this work, we use the MI methodology to determine the interfacial free energy between the \ch{CO2} hydrate crystalline solid and the aqueous solution of \ch{CO2} at coexistence conditions. According to this, the work needed to form a thin crystalline slab of hydrate in the aqueous solution, $\Delta G^{hw}$, is directly related with the hydrate-water interfacial free energy, $\gamma_{hw}$, through its definition,

\begin{equation}
\gamma_{hw}=\dfrac{\Delta G^{hw}}{2\mathcal{A}}
\label{gamma_hw1}
\end{equation}

\noindent
Here $\mathcal{A}$ is the area of the interface between the hydrate and the aqueous solution of \ch{CO2} and the factor 2 arises because once the crystalline solid slab is created there are two hydrate--water interfaces. It is important to remark that Eq.~(\ref{gamma_hw1}) only holds if $\Delta G^{hw}$, the reversible work needed to induce the formation of a hydrate slab, is determined at coexistence conditions. 
These means that all phases in contact have the same chemical potential, i.e., the \ch{CO2} hydrate solid phase and the two
liquid phases coexist at the same temperature and pressure.
%\sout{According to the MI methodology (see the Supporting Information (SI) for a detailed
%account of it), the formation of the hydrate crystal slab is induced placing attractive %interaction sites in the \ch{H2O}-rich liquid phase at the equilibrium positions of the %oxygen atoms of water in one of principal planes of the sI structure of the \ch{CO2} %hydrate.}

Homogeneous nucleation of any crystalline solid phase from its liquid, including hydrates, is an activated process~\cite{Debenedetti1996a,Sanz2013a,Espinosa2014a}.
Typically, times and lengths at which nucleation occurs are in the nanoseconds and in the nanometers time and length
scales, respectively~\cite{Sloan2008a,Debenedetti1996a}. This is why homogeneous nucleation is considered a rare event, which means that it is nearly
impossible to observe spontaneously, experimentally or by means of molecular simulation, the growth of a solid phase from
its liquid~\cite{Sloan2008a,Debenedetti1996a}.

To overcome this problem and induce the formation of the hydrate crystal slab, we place attractive interaction
sites in the \ch{H2O}-rich liquid phase at the equilibrium positions of the oxygen atoms of water in one of principal planes of the sI
structure of the \ch{CO2} hydrate. The particular number and positions occupied by the attractive interaction sites are described in
the next Section. The most appropriate intermolecular potential to act as catalyst of the hydrate slab is a continuous version
of the square-well potential~\cite{Espinosa2014a} characterized by two molecular parameters: the range of interaction, $r_{w}$, and the well depth, $\varepsilon$. See
the works of Espinosa \textit{et al.}~\cite{Espinosa2016a,Espinosa2014a,Espinosa2015a} for further details on the particular implementation of this continuous square-well
intermolecular potential. This mold of attractive sites provokes attractive interactions between them and water molecules. In
other words, when the mold is switched off, water molecules can diffuse freely in the fluid. However, as the mold is
progressively switched on (increasing the well depth, $\varepsilon$, from 0 to a maximum value, $\varepsilon_{m}$), an increasing number of sites of the
mold are occupied by water molecules. If $r_{w}$, the attractive interaction range of the wells, is sufficiently narrow a \ch{CO2} hydrate
crystal slab is formed around the positions of the mold wells. Note that this parameter must be small enough to
accommodate only one water molecule when the wells are switched on. According to
the Mold Integration methodology, the interfacial free energy
depends on the particular value of the well radius, $r_{w}$, used during the
calculations. Consequently, it is necessary to calculate \emph{a priori} which
value gives the correct estimate for $\gamma_{hw}$. This value is known as optimal well radius
and is usually refer to as $r_{w}^{0}$. If $r_{w}<r_{w}^{0}$, the free energy
of the hydrate phase is lower than that of the aqueous solution of CO$_{2}$,
and the system evolves quickly to the equilibrium solid phase through a first-order
phase transition. In other words, the system freezes completely. Note that this
is not the desired situation since the method is based on the calculation of the
reversible work needed to induce the formation of a thin hydrate slab.
Since $\gamma_{hw}$ is calculated using a thermodynamic integration
when the mold is progressively switched on, only reversible paths are allowed,
i.e., the Mold Integration must be used for values $r_{w}>r_{w}^{0}$. According to
Espinosa \emph{et al.}, it is not advisable to perform thermodynamic integration
for $r_{w}=r_{w}^{0}$, the state at which the free energies of the hydrate
and aqueous solution phases are equal. Hence, several calculations of
$\gamma_{hw}$, for different $r_{w}$ values are performed (with $r_{w}>r_{w}^{0}$).
The correct value of $\gamma_{hw}$, at coexistence, is then calculated
extrapolating the function $\gamma_{hw}=\gamma_{hw}(r_{w})$ to $r_{w}^{0}$. For further
details, we recommend the original manuscript of Espinosa \emph{et al.}~\cite{Espinosa2014a} (particularly,
the section explaining the Figure 2 of the paper). This methodology has been also used successfully by these authors to determine accurately the interfacial free energy of the Lennard-Jones and hard-sphere
systems~\cite{Espinosa2014a}, sodium chloride~\cite{Espinosa2015a}, and several models of water~\cite{Espinosa2016a} at coexistence conditions.

The free energy difference between the fluid and the fluid with the mold fully occupied, $\Delta G_{m}$, can be obtained using a thermodynamic integration along a path in which the attractive interaction between the mold sites and water molecules, $\varepsilon$, is gradually switched from
the state in which the interaction is switched off, $\varepsilon = 0$ (all water molecules in fluid phase) to another in which the mold sites are fully switched on, $\varepsilon=\varepsilon_{m}$ (some water molecules form a hydrate
slab). $\Delta G_{m}$ can be written as,~\cite{Espinosa2016a}.

\begin{equation}
\Delta G_{m}= - \int_{0}^{\varepsilon_{m}} \langle N_{fw}(\varepsilon) \rangle_{_{NP_{z}\mathcal{A}T}}\, d \varepsilon
\label{deltaG_m}
\end{equation}

\noindent
where $\langle N_{fw}(\varepsilon) \rangle_{_{NP_{z}\mathcal{A}T}}$ is the number of filled wells averaged over the configurations generated using the isothermal-isobaric or $NP_{z}AT$ ensemble when the interaction energy is $\varepsilon$. Here $P_{z}$ and $T$
are the coexistence pressure and temperature, respectively. To determine the number of sites of the mold that are filled by water molecules we define
a new class of atoms in GROMACS representing the associating sites of the mold. The interaction
energy between the wells and the oxygen atoms of water molecules is
set equal to $-\varepsilon$ and $0$ with the rest of atoms in the system. The fluid-well
interaction energy $U_{fw}(\varepsilon)$, for a given value of $\varepsilon$, is then
calculated. $U_{fw}$ is proportional to $-\varepsilon$ since each water molecule
can only occupy one well site ($r_{w}$ is chosen with this condition; see below for
further details). Consequently, the number of sites of the mold that are filled by
water molecules are calculate as $N_{fw}=U_{fw}/(-\varepsilon)$.
Particularly, we have checked that $U_{fw}\approx -448\,k_{B}T$ when $\varepsilon\rightarrow\varepsilon_{m}$
($\varepsilon_{m}=8\,k_{B}T$) in all cases, which implies that $N_{fw}\approx 56$. This numerical
calculation is confirmed by a ``visual'' inspection of the system making movies of the
evolution of the system and using VMD software to follow the evolution of the water -- well interactions. $\Delta G^{hw}$ is calculated subtracting the contribution that represents the interaction energy between the water molecules and the mold, $-N_{w}\varepsilon_{m}$.
 
\begin{equation}
\Delta G^{hw}= N_{w}\varepsilon_{m} + \Delta G_{m}
\label{gamma_hw2}
\end{equation}

%\sout{See the SI for a detailed account of the methodology and its
%extension to deal with \ch{CO2} hydrates.}

\noindent
Here $N_{w}$ is the total number of wells. The contribution $-N_{w}\varepsilon_{m}$ represents the interaction energy between the water molecules and
the mold and must be subtracted from $\Delta G_{m}$ to obtain the reversible work to form the crystalline slab~\cite{Espinosa2015a}. The value of $\varepsilon_{m}$ must satisfy
two conditions: (1) it should be high enough to guarantee that every well of the mold is occupied by a water molecule in the
final state of the integration path; and (2) $\varepsilon_{m}$ should be low enough to ensure that integrand of Eq. \eqref{deltaG_m} behaves smoothly when $\varepsilon$
varies from 0 to $\varepsilon_{m}$. Large values of $\varepsilon_{m}$ produce large variations of $\langle N_{fw}(\varepsilon) \rangle_{_{NP_{z}\mathcal{A}T}}$ and a large number of intermediate states must be used to ensure an accurate determination of the interfacial free energy. In this work, we have used $\varepsilon_{m} = 8\,\text{k$_{B}$T} $, where k$_{B}$ is the Boltzmann constant and T the temperature of the system. This value allows to ensure that all sites of the mold are filled by water molecules for values of $\varepsilon < \varepsilon_{m}$~\cite{Espinosa2014a,Espinosa2016a}.

The calculation of $\Delta G^{hw}$ following Eqs.~\eqref{deltaG_m} and \eqref{gamma_hw2} can be used to determine the \ch{CO2} hydrate-water interfacial free energy only
if it represents the reversible work to induce the formation of a hydrate slab. Hence, one must ensure that the system does
not cross a first-order phase transition along the integration path. In other words, in all the states chosen to perform the
thermodynamics integration the system cannot fully form a solid crystalline (hydrate) phase. According to Espinosa \textit{et al.}~\cite{Espinosa2015a},
this is practically done using potential ranges, $r_{w}$, wider than a certain optimal range value, $r_{w}^{0}$. A full explanation on how $r_{w}^{0}$ is
determined in this work is explained in the Results and discussion Section.

\textbf{Simulation details.} The method for calculating the SF interfacial free energy
involves four steps: (1) preparation of the simulation box with the fluid phases that coexist at thermodynamic conditions at which
the interfacial free energy is obtained; (2) determination of the
optimal well radius $r_{w}^{0}$; (3) calculation of the SF interfacial free energy as a function of the well radius $r_{w}$ using thermodynamic integration; and (4) extrapolation of the interfacial free energy to $r_{w}^{0}$. In this section, we concentrate in step (1). Steps (2)-(4) are
presented in the next Sections.

\begin{figure}[h!]
\centering
\includegraphics[angle=0,width=1.10\columnwidth]{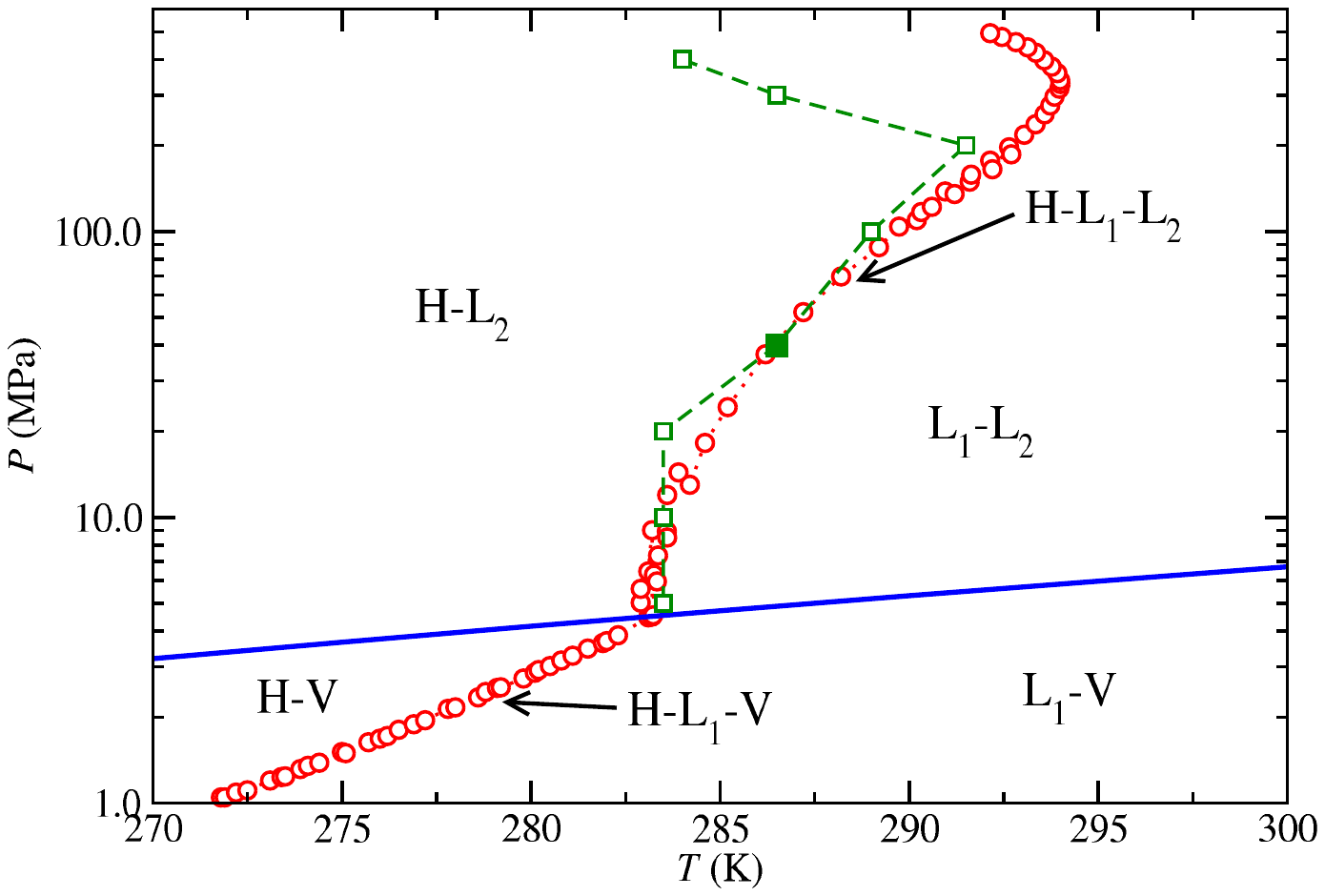}
\caption{\small Pressure-temperature projection of the three-phase coexistence line (\ch{CO2} hydrate -- \ch{H2O} -- \ch{CO2}) of the \ch{CO2} hydrate. Red circles represent the experimental data taken from the literature~\cite{Nakano1998a}, the blue curve the experimental vapor pressure of pure \ch{CO2}~\cite{Lemmon2019a}, and the green squares the results obtained by M\'{\i}guez \textit{et al.}~\cite{Miguez2015a} from molecular dynamics simulation using
the direct coexistence technique. H corresponds to the hydrate phase, L$_{1}$ and L$_{2}$ represent the \ch{CO2}-rich and
\ch{H2O}-rich liquid phases, respectively, and V is the vapor.
The filled green square represents the state point ($287\,\text{K}$ and $40\,\text{MPa}$) at which simulations are performed in this work.}
\label{pt}
\end{figure}

Contrary to what happens in pure systems~\cite{Espinosa2016a,Espinosa2014a}, the use of the
MI technique for hydrates requires special attention because the
coexistence conditions of the hydrate-water interface involves two
different components (\ch{H2O} and \ch{CO2}) and three phases in equilibrium, the \ch{CO2} hydrate solid, \ch{H2O}-rich liquid, and \ch{CO2}-rich liquid phases. We use the phase diagram of the \ch{CO2} hydrate recently determined by some of us~\cite{Miguez2015a}, combining the well-known direct coexistence technique, the TIP4P/Ice model for water~\cite{Abascal2005b}, and the TraPPE model for \ch{CO2}~\cite{Potoff2001a}, to select the conditions at which all simulations are performed in this work: $40\,\text{MPa}$ and $287\,\text{K}$. According to Figure~\ref{pt}, these conditions (filled green square) correspond to a state point along the dissociation line of the \ch{CO2} hydrate. Note that the vapor phase only exists at temperatures and pressure below the vapor-liquid coexistence curve of pure \ch{CO2} (blue continuous curve). 

It is important to recall here that we use, not only the same molecular models for
water (TIP4P/ice) and CO$_{2}$ (TraPPE), but also the same modified Lorentz-Berthelot combining rules. According
to the previous work of M\'{\i}guez \emph{et al.}~\cite{Miguez2015a}, we assume that the unlike interaction parameters are given
by the modified Lorentz-Berthelot combining rules,

\begin{equation}
        \sigma_{ij}=\dfrac{1}{2}(\sigma_{ii}+\sigma_{jj})\tag{4}
\end{equation}

\noindent
and

\begin{equation}
        \epsilon_{ij}=\chi(\epsilon_{ii}\,\epsilon_{jj})^{1/2}\tag{5}
\end{equation}

\noindent
Here $\chi=1.13$. The use of this
positive deviation of the Berthelot rule, $\chi=1.13$, ensures an
accurate description of the experimental dissociation line up to $2000\,\text{bar}$~\cite{Miguez2015a}. Particularly,
this allows to predict very accurately the coexistence temperature
at $400\,\text{bar}$.

Although there are experimental values~\cite{Nakano1998a, Lemmon2019a} of the precise location
of the three-phase line at a wide range of pressures, we only
concentrate in this work on the interfacial free energy of the \ch{CO2}
hydrate-water interface at $40\,\text{MPa}$ and $287\,\text{K}$. These conditions are selected because agreement between experimental data taken from the literature and computer simulation predictions for coexistence is excellent. In addition to that, and according to the available experimental data from Uchida~\cite{Uchida1999a,Uchida2002a} and Anderson~\cite{Anderson2003a,Anderson2003b},
$\gamma_{hw}$ is independent of temperature and pressure along the three-phase dissociation line of \ch{CO2} hydrates.

We perform MD simulations in combination with the direct coexistence technique in the isothermal-isobaric or $NP_{z}\mathcal{A}T$ ensemble~\cite{Allen2017a,Frenkel2002a}. We use a parallelepiped simulation box of volume $V=L_{x}\times L_{y}\times L_{z}$, where $L_{x}$, $L_{y}$, and $L_{z}$ are the dimensions of the simulation box. $L_{x}$ and $L_{y}$ are kept constant and only $L_{z}$ is varied along the simulation. This ensures that system is under the equilibrium normal pressure (perpendicular to the hydrate slab formed when the mold is switched on). It also avoids stress in the system, keeping $L_{x}$ and $L_{y}$ constants and consistent with the equilibrium unit cell of the hydrate phase at coexistence conditions.

To prepare the simulation box used in this work, we first consider two homogeneous liquid systems, a pure water liquid phase formed by 736 molecules of water and two pure \ch{CO2} liquid phases formed by 128 \ch{CO2} molecules each of them. Dimensions of the initial simulation boxes are chosen to be consistent with the size of the hydrate unit cell along the $x-$ and $y-$axis that will be formed when the mold is switched on. According to this, $L_{x}=L_{y}=24\,\text{\AA}$ and kept fixed during all the simulations. The liquid-liquid (LL) system is equilibrated along 5 ns. After this, the equilibrium value of the length of the simulation box along the $z-$axis was $L_{z}\approx 66.5 \,\text{\AA}$. The size of the simulation box and the number of molecules used here are consistent with parameters used by
M\'{\i}guez \emph{et al.} to determine the dissociation line of the CO$_{2}$ hydrate, and consequently,
with the temperature at which the hydrate is in equilibrium with the water-rich and CO$_{2}$-rich
liquid phases at $400\,\text{bar}$. Particularly, the lateral size of the simulation box fits
with a $2\times2$ unit cells of the hydrate, consistent with the size of the hydrate phase used
by M\'{\i}guez and coworkers to determine the three-phase line ($2\times2\times2$ unit cells). The size
used in our work is also in agreement with the lateral sizes previously used by Vega and collaborators
to determine the interfacial free energy of Lennard-Jones and pseudo hard-sphere systems~\cite{Espinosa2014a},
several models of water~\cite{Espinosa2016a}, and NaCl~\cite{Espinosa2015a} from the MI technique.

The equilibrium configurations corresponding to the final simulation boxes
are formed from a H$_{2}$O-rich liquid slab of $\sim32\,\text{\AA}$, a CO$_{2}$-rich liquid slab
of $\sim20\,\text{\AA}$, and a thin hydrate slab of $\lesssim 10\,\text{\AA}$. Note that the
hydrate slab is induced by the associating sites of the mold when are switched
on ($\varepsilon\rightarrow\varepsilon_{m}$). The typical size of a simulation box along the $z-$axis
and the width of the different phases used in our work are similar to those used previously by Vega and 
coworkers for the methane hydrate~\cite{Conde2010b,Conde2013a}, which simulate boxes of $24\times24\times55\,\text{\AA}^{3}$ and
$24\times24\times67\,\text{\AA}^{3}$ (in this case, $\sim22\,\text{\AA}$ for the
H$_{2}$O-rich liquid slab and $\sim21\,\text{\AA}$ for the methane-rich fluid slab),
and by Fern\'andez-Fern\'andez for the methane hydrate
in oceanic conditions~\cite{Fernandez-Fernandez2019a} and fluorinated methane derivatives
hydrates~\cite{Fernandez-Fernandez2021a}, which simulate
simulation boxes of size $24\times24\times72\,\text{\AA}^{3}$. The current work must
be understood as a first attempt to
determine, from first principles, the interfacial free energy of a hydrate at coexistence
conditions. However, we think that finite-size effects should play an important
role, not only in the final value of the interfacial energy but also in the precise location
of the three-phase coexistence line of the hydrate. Unfortunately, this is out of the scope
of this work.

We also place $N_{w}=56$ attractive interaction sites at the equilibrium positions of the oxygen atoms of water located at one of the principal planes of the sI structure of the \ch{CO2} hydrate. We recommend to see Figure 1 of our previous work~\cite{Miguez2015a} to identify the particular positions of the oxygen atoms at which the mold sites are placed. Note that the sites are switched off during the setup of the simulation box. They are only switched on when the MI method is used, as it is explained in the next section. After this, the final simulation box is built linking up the three bulk boxes forming a \ch{CO2}-rich liquid - \ch{H2O}-rich liquid - \ch{CO2}-rich liquid or L$_{1}$ -- L$_{2}$ -- L$_{1}$ system. It is important to note that two issues at this point. Firstly, the presence of the
CO$_{2}$-rich liquid slab in the simulation box is necessary to ensure that calculations
are performed at equilibrium coexistence conditions between the water-rich,
the CO$_{2}$-rich, and the thin hydrate slab formed when the sites of the mold are switched on. Note
that when the hydrate slab grows, CO$_{2}$ molecules in the H$_{2}$O-rich liquid phase
are incorporated into the solid phase. This produces a decreasing of the equilibrium
concentration of the CO$_{2}$ in the aqueous solution that is compensated incorporating
extra CO$_{2}$ molecules from the CO$_{2}$-rich liquid reservoir. Since this non-aqueous
phase is essentially a pure phase of carbon dioxide (solubility of CO$_{2}$ in water
is completely negligible), chemical equilibrium between the three phases is guaranteed. Secondly, the CO$_{2}$-rich liquid phase is never in
direct contact with the hydrate slab. This solid phase induced by the mold is always in
contact with the aqueous solution of CO$_{2}$. However, since the solubility of CO$_{2}$ in water is
very small, the H$_{2}$O-rich liquid phase is usually named in the literature
water phase. For this reason, and following the nomenclature used in the literature,
the interfacial energy in this work is referred to as the interfacial free energy
of the CO$_{2}$ hydrate -- water interface.

We use a Verlet leapfrog algorithm~\cite{Cuendet2007a}  with a time step of 0.0015 ns to solve the Newton's equations. All simulations are run at constant temperature and pressure using a Nos\'e-Hoover thermostat~\cite{Nose1984a} and a Parrinello-Rahman barostat~\cite{Parrinello1981a}. Relaxation times used in the thermostat and barostat are 2 and 1 ps, respectively. Long-range interactions due to electrostatic interactions are determined using the Particle Mesh Ewald technique~\cite{Essmann1995a}. We use a cutoff radius for both dispersive interactions and the real part of electrostatic interactions of $10\,\text{\AA}$. The final simulation box obtained as explained above has been used as initial simulation box in the rest of the work to determine the optimal value of the attractive interaction range of the wells, $r_{w}^{0}$, and to calculate the number of filled wells for different values of $r_{w}$ and $\varepsilon$.

%\sout{The initial simulation box, prepared following a standard procedure, is formed by %linking up three bulk boxes forming a \ch{CO2}-rich liquid --
%\ch{H2O}-rich liquid -- \ch{CO2}-rich liquid or L$_{1}$ -- L$_{2}$ -- L$_{1}$ system. This %simulation box is used in the rest of the work to determine the optimal value of the %attractive interaction range of the wells, $r_{w}^{0}$, and to calculate the number of filled %wells for different values of $r_{w}$ and $\varepsilon$. The rest of the simulation details %are provided in the SI.}

\section{Results and discussion}

\textbf{Determination of $r_{w}^{0}$.} We have checked, following the works of Espinosa \textit{et al.}~\cite{Espinosa2016a,Espinosa2014a,Espinosa2015a}, that it is possible to induce a \ch{CO2} hydrate slab using a mold with appropriate values of $r_{w}$ and $\varepsilon$, the well radius and depth of the mold, respectively. As it is discussed previously, $\varepsilon$ values in the range $0<\varepsilon\le\varepsilon_{m}$, with $\varepsilon_{m}=8k_{B}T$, once the mold is switched on, allow the water molecules to occupy all the sites of the mold, promoting the growth of the \ch{CO2} hydrate slab. To check this, we have used first a mold of attractive sites with a very low value of $r_{w}=0.760\,\text{\AA}$ and $\varepsilon=\varepsilon_{m}=8k_{B}T$. As we will see later in this section, very low values of $r_{w}$ are not appropriate for calculating interfacial free energies. However, this election of parameters is useful at this level for checking if the mold is really able to induce the formation of a \ch{CO2} hydrate phase. 

\begin{figure}[h!]
\centering
\includegraphics[angle=0,width=\columnwidth]{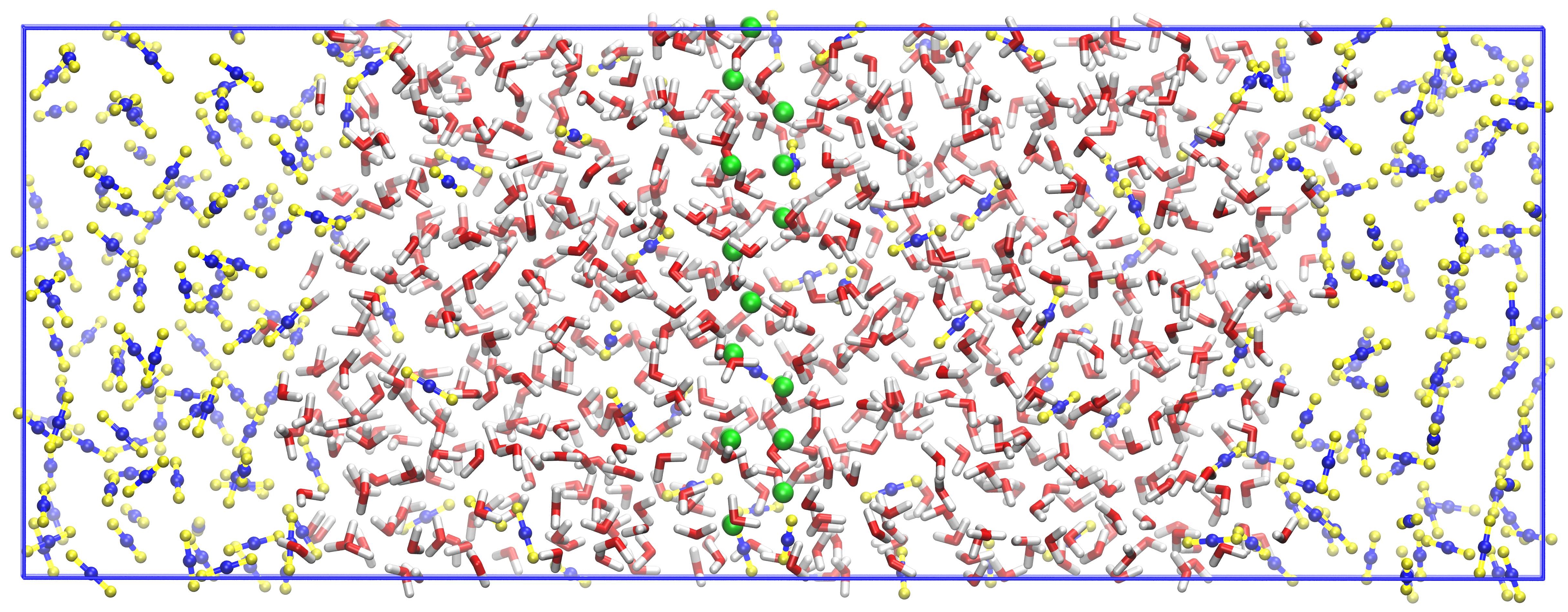}
\includegraphics[angle=0,width=\columnwidth]{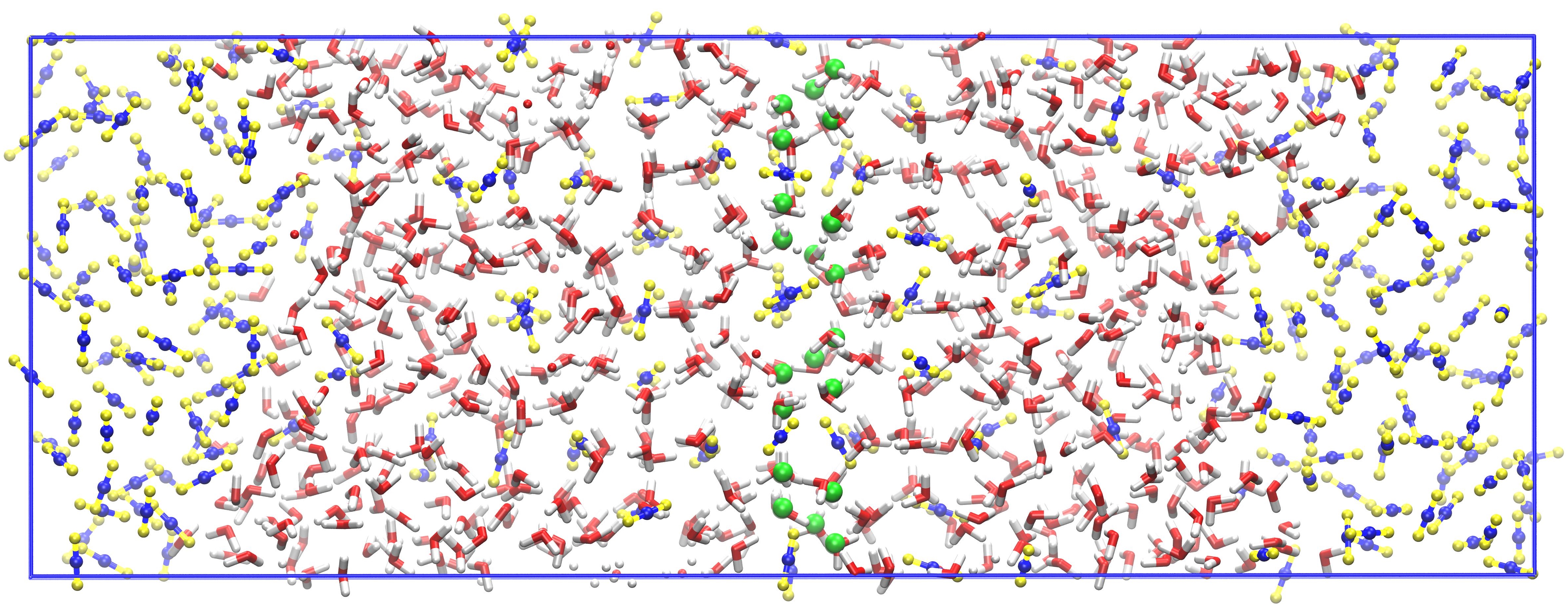}
\includegraphics[angle=0,width=\columnwidth]{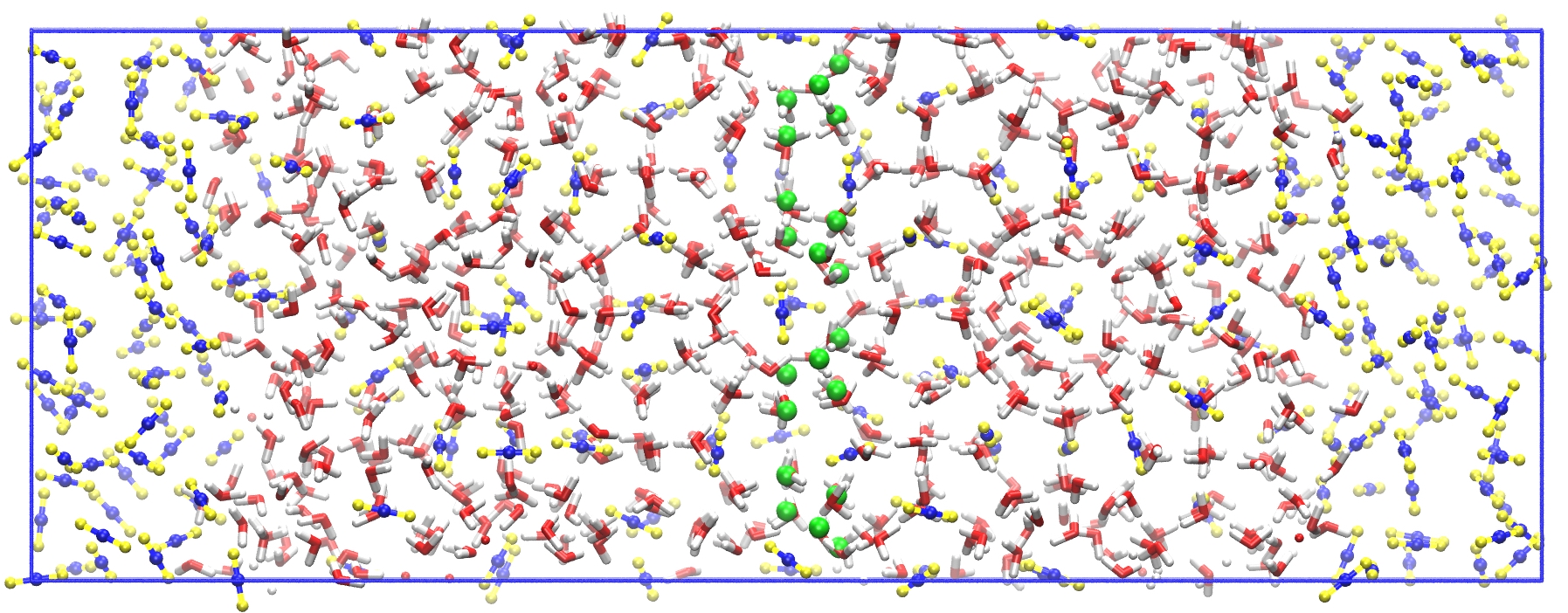}
\caption{\small Snapshots showing the crystallization of the
\ch{CO2} hydrate phase from the water-\ch{CO2} two-phase coexistence
at $40\,\text{MPa}$ and $287\,\text{K}$. Different parts show the simulation box for (top) $t=0$, (middle) $t=50\,\text{ns}$, and
(bottom) $t=200\,\text{ns}$. Red and white licorice representation corresponds to oxygen and hydrogen atoms of water, respectively, yellow and blue spheres (Van der Waals representation) correspond to carbon and oxygen atoms of \ch{CO2}, respectively, and green spheres (Van der Waals representation) correspond to the mold attractive sites with $r_{w}=0.760\,\text{\AA}$ and $\varepsilon=8\,k_{B}T$.} 
\label{snapshot}
\end{figure}

Figure~\ref{snapshot} shows three snapshots extracted from MD trajectories of the simulation prepared as explained in the previous Section (LL interfacial system for the \ch{CO2} + \ch{H2O} system), with the mold switched on ($r_{w}=0.760\,\text{\AA}$ and $\varepsilon=8\, k_{B}T$). As can be seen, at $t=0\,\text{ns}$ (top) the system exhibits LL immiscibility, with a \ch{H2O}-rich liquid phase (center of the simulation box) in coexistence with a \ch{CO2}-rich liquid phase (left and right of the simulation box). The attractive sites of the mold (green spheres) occupy the equilibrium positions of the oxygen atoms of water of one lattice plane of the \ch{CO2} hydrate. $50\,\text{ns}$ later (middle), water molecules have arranged around the equilibrium crystalline positions and \ch{CO2} molecules are trapped inside the hydrate cages, forming 4-5 hydrate layers occupied by the guest molecules. Note that there are still water molecules in the aqueous phase between the incipient forming hydrate solid phase and the \ch{CO2}-rich liquid phases located at the borders of the simulation box. Finally, at $t=200\,\text{ns}$ (bottom) all water molecules are in the hydrate solid phase in coexistence with the \ch{CO2}-rich liquid phase. Note that configurations shown in the middle and bottom parts of the Figure are never considered
during the evaluation of the interfacial free energy
of the hydrate. These configurations are only presented here to demonstrate that the mold
used in this work is able to induce a CO$_{2}$ hydrate slab,
and also, that using a small value of $r_{w}$ ($0.760\,\text{\AA}$) the hydrate eventually
grows and occupies nearly the whole simulation box. It will be clear later in this section that the mold used here is not appropriate to calculate interfacial energies. However, these results show that the MI method can be used to induce the formation of hydrates in the corresponding aqueous solution.
The MI methodology allows to calculate the interfacial free energy
of the \ch{CO2}- hydrate-water system as a function of $r_{w}$. Although, it is necessary to determine the optimal value of $r_{w}$, denoted by $r^{0}_{w}$, that gives the right value of $\gamma_{hw}$. This election is key in this
technique since the uncertainty of $r^{0}_{w}$ is the main source of error in the calculation of the interfacial free energy~\cite{Espinosa2016a,Espinosa2014a}. It is possible to select the optimal value $r^{0}_{w}$ from the inspection of $n_{h}$, the
number of molecules of the crystalline solid slab formed when the
mold is switched on.

We perform simulations in the $NP_{z}\mathcal{A}T$ ensemble starting from the equilibrium LL initial configuration previously generated, with the mold fully switched on ($\varepsilon=\varepsilon_{m}$), for different values of $r_{w}$. For each selected $r_{w}$, we perform several independent trajectories to obtain a statistical picture of the behavior of the system. The use of the MI technique for the determination of the \ch{CO2} hydrate-water interfacial free energy requires longer simulation times compared with other systems~\cite{Espinosa2016a,Espinosa2014a,Espinosa2015a}, not only because one has to start from a two-phase coexisting system, but also due to the nature itself of hydrates. The formation of the solid crystalline slab involves the arrangement of water molecules around the equilibrium crystalline positions catalyzed by the mold when is switched on. Note that the sites of the mold are located in the middle of the simulation box (\ch{H2O}-rich liquid phase) and the arrangement of water molecules is usually a quick process compared with the full formation of the hydrate slab. However, the \ch{CO2} molecules must also occupy their equilibrium positions in order to form a stable hydrate slab. Since solubility of \ch{CO2} in \ch{H2O} at the coexistence conditions is very low, there are not enough \ch{CO2} molecules in the \ch{H2O}-rich liquid phase to form a stable hydrate slab. Consequently, it is necessary that \ch{CO2} molecules diffuse from the sides of the simulation box to the center of it in order a stable hydrate slab will be formed. Note that the thermodynamic stability of hydrate phases is achieved due to the presence of the guest molecules inside the hydrate crystalline structure~\cite{Sloan2008a}. According to this, simulation lengths considered in this work have to be much longer than in previous studies. For example, Espinosa \textit{et al.}~\cite{Espinosa2016a,Espinosa2014a} analyze the behavior of $n_{h}$ for Lennard-Jones and water systems during $1\,\text{ns}$, approximately, whereas in this work we monitor the number of water molecules in the hydrate crystalline slab along $100\,\text{ns}$, two orders of magnitude larger. Fortunately, these
simulations are feasible thanks to the optimized software packages, such as GROMACS, and the new hardware computational resources available now-a-days.

According to Vega and collaborators~\cite{Espinosa2016a,Espinosa2014a}, if $r_{w}< r_{w}^{0}$, a hydrate crystalline slab should grow in all trajectories considered. In particular, evolution of trajectories with time do not show any induction period. In other words, the number of water molecules in the crystal cluster grows very quickly, indicating that there is no free energy barrier between the fluid and the solid, and therefore the formation of the hydrate slab takes place spontaneously. Contrary, if $r_{w}> r_{w}^{0}$, there exists a free energy barrier between the fluid and the solid that must be overcome. This can be observed in the evolution of the number of water molecules in the crystal clusters (hydrate) with time. Some of the trajectories show an induction period before the slab grows, i.e., the number of water molecules in the crystal cluster does not grow during a time period. It is possible that some trajectories do not show slab formation, especially if the $r_{w}$ values are very high compared with those of $r_{w}^{0}$.

The key magnitude in the previous discussion is $n_{h}$, the number of water molecules in the crystalline slab formed when the mold is switched on. To determine this magnitude, we use the rotationally invariant local bond order parameters, $\overline{q}_{i}$, proposed by Lechner and Dellago~\cite{Lechner2008a} (see Eqs.~(1), (5) and (6) in their work). 
These order parameters are similar to the well-known traslational
and orientational order parameters proposed by Truskett \textit{et al.}~\cite{truskett2000} and Chau and Hardwick~\cite{chau1998}, respectively, that have been recently used by Sebastiani \textit{et al.}~\cite{sebastiani2020} to characterize the “phase” of water within supramolecular tetrahedral cages. This approach or similar techniques have been previously used by several authors in the literature to identify if a molecule is a fluid-like or solid-like particle~\cite{vanDuijneveldt1992a, tenWolde1995a, Lynden-Bell1993a, tenWolde1996a, tenWolde1997a}, including Espinosa and coworkers \cite{Espinosa2016a, Sanz2013a}. In this work, we use a similar approach than Sanz \textit{et al.}~\cite{Sanz2013a}, that analyzed the crystallization of water molecules in ice Ih solid phase but with two important differences: (1) we use a $\overline{q}_{3}-\overline{q}_{6}$ representation, instead of the
usual $\overline{q}_{4}-\overline{q}_{6}$ proposed by Lechner and Dellago and other authors; and (2) the cutoff distance to identify neighbors for the calculation of $\overline{q}_{i}$ between the oxygen atoms was chosen in a different way. 

\begin{figure}[h!]
\centering
\includegraphics[angle=0,width=1.10\columnwidth]{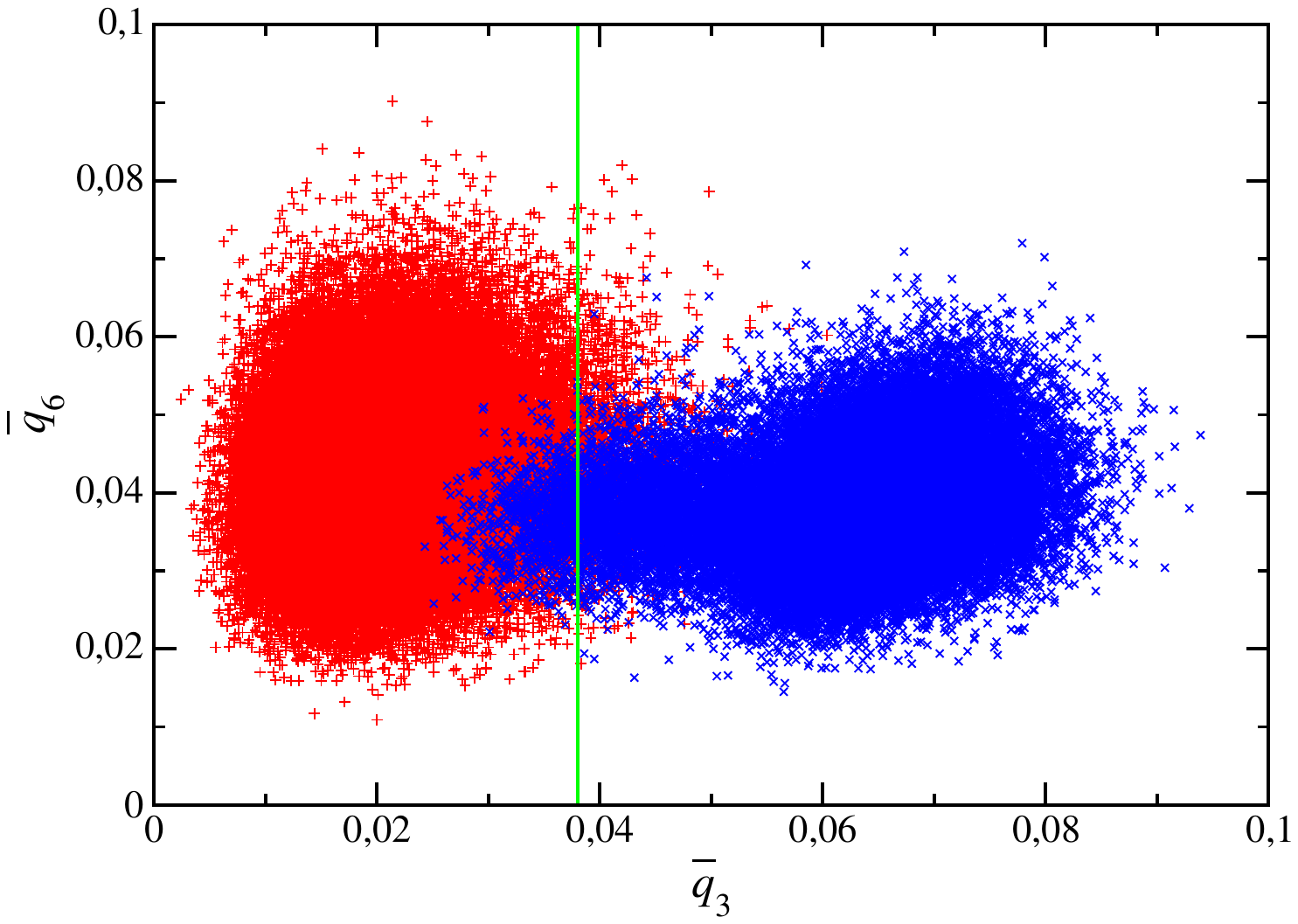}
\caption{\small Values of $\overline{q}_{3}$ and $\overline{q}_{6}$ for
water molecules in a system formed by 736 water molecules and 256 CO$_{2}$ molecules at coexistence conditions ($40\,\text{MPa}$ and $278\,\text{K}$) corresponding to $2.5\times10^{7}$ MD steps of simulation ($50\,\text{ns}$). Blue crosses represent water molecules in the liquid phase and red pluses to water molecules in the hydrate phase.}
\label{q3_q6}
\end{figure}

The combination of $\overline{q}_{4}$ and $\overline{q}_{6}$ bond-order
parameters does not allow to discriminate between solid-like and liquid-like molecules for systems in which stable solid phase is a hydrate~\cite{Reinhardt2012a}. In the particular case of sI hydrates, it is possible to use a threshold value $\overline{q}_{3,t}\approx 0.038$ in the $\overline{q}_{3}-\overline{q}_{6}$ representation
that separates the liquid from the hydrate as it is shown in Figure~\ref{q3_q6}. This value has been obtained from the percentage of mislabeled particles as a function of the threshold value chosen for the $\overline{q}_{3}$ parameter~\cite{Sanz2013a}. On the other hand, Sanz \textit{et al.}~\cite{Sanz2013a} used a cutoff distance for calculating the $\overline{q}_{i}$ parameters of $3.5\,\text{\AA}$, that corresponds to the position of the first minimum of the oxygen-oxygen pair correlation function in the liquid phase. However, we use a value of $5.5\,\text{\AA}$, that corresponds to the position of the second minimum of the oxygen-oxygen pair correlation function in the hydrate phase. This allows, together with the $\overline{q}_{3}-\overline{q}_{6}$ representation, to distinguish clearly between solid-like and liquid-like molecules for \ch{CO2} hydrates.

\begin{figure*}[h!]
\centering
\includegraphics[angle=0,width=\columnwidth]{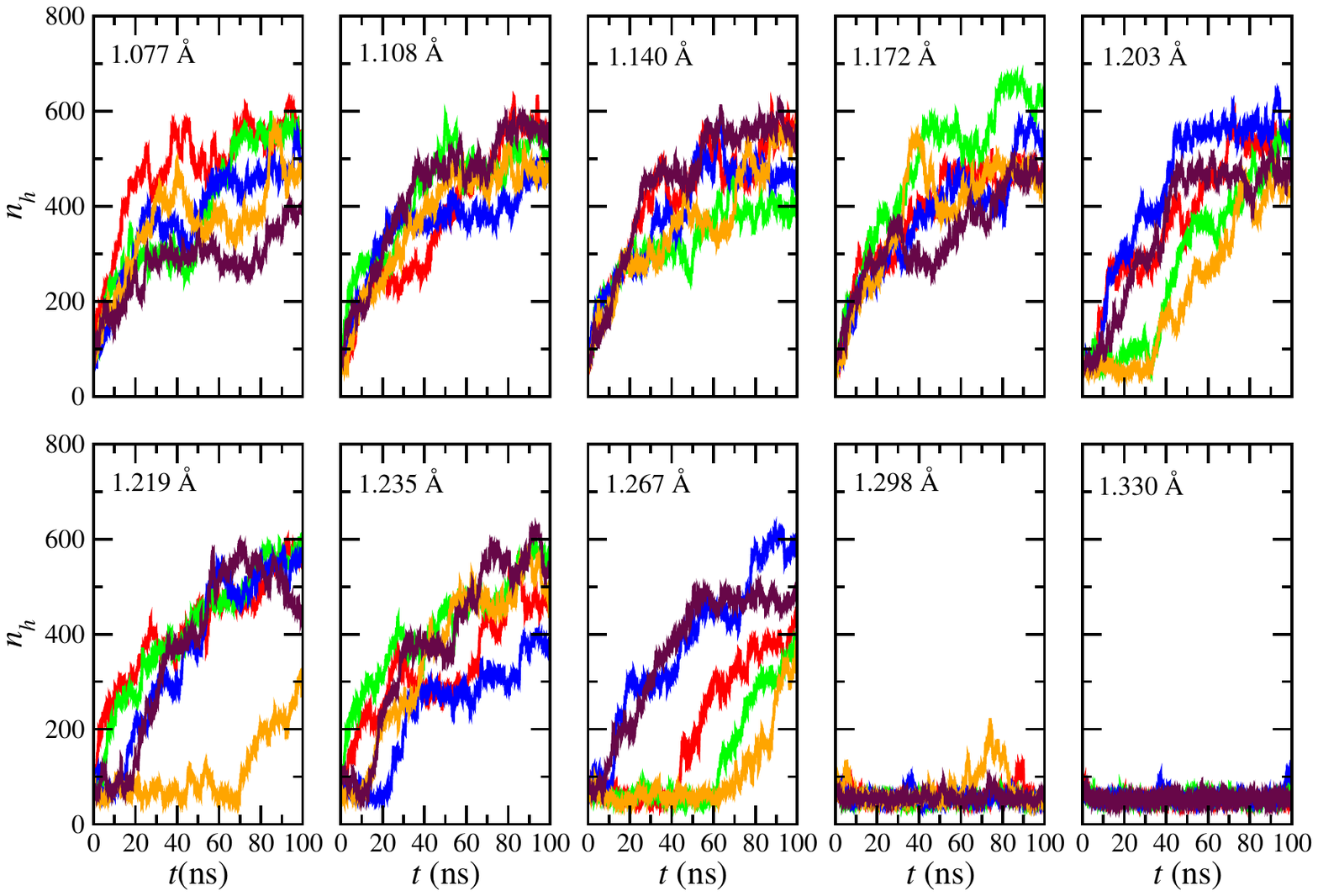}
\caption{\small Number of water molecules in the crystal slab, $n_{h}$, as a function of time for several trajectories and different well radio, $r_{w}$ (as indicated in the legend). All
simulations are performed at coexistence conditions ($40\,\text{MPa}$ and $287\,\text{K}$). In all cases, $\varepsilon = 8\,k_{B}T$. Each color represents an independent trajectory generated using different seeds starting from the same fluid configuration.}
\label{nh_vs_t}
\end{figure*}

Using this information, we have analyzed the time evolution of
$n_{h}$, the number of water molecules in the \ch{CO2} hydrate slab, for several well radius considered in this work. In particular, we have studied eighteen values ranging from $0.76\,\text{\AA}$ (0.24$\sigma$) to $1.457\,\text{\AA}$ (0.46$\sigma$), with $\sigma=3.1668\,\text{\AA}$ the diameter associated with the Lennard-Jones intermolecular potential of the TIP4P/ice water model~\cite{Abascal2005b}.
Selected results for $n_{h}$, as a function of time, are shown in Figure~\ref{nh_vs_t}. The complete set of results are presented in the SI material. As we have mentioned previously, the key point to
determine $r_{w}^{0}$ is the identification of what Espinosa \textit{et al.}~\cite{Espinosa2016a} call induction period along the evolution of the number of water molecules in the crystal slab $n_{h}$, as a function of time. According to their seminal work, the induction period is the time interval observed from the beginning of a simulation from a fluid configuration (t = 0 ns) up to a time in which $n_{h}$ starts to grow, on average, monotonically. During the induction period $n_{h}$ fluctuates around its initial value. As we have already mentioned, the existence of an induction period indicates that there is a free energy barrier between the fluid and solid states that avoids the formation of a crystal slab. Contrary, the absence of such an induction period indicates that there is no free energy barrier.

Following the work of Espinosa \textit{et al.}, for each $r_{w}$ value we
have run five independent trajectories using different seeds
starting from the same fluid configuration. This allows to have a
statistical picture of the behavior of the system upon switching the mold on. We have paid attention to the effect of different initial configurations and different
initial seeds on the evolution of $n_{h}$. In all cases, the general trend
of $n_{h}=n_{h}(t)$ was qualitatively identical to that shown here. In particular, we have
found that for a fixed $r_{w}$ value, the set
of the trajectories obtained follow a path that can be classified in one of scenarios
described in the following paragraph. Each trajectory is run over 100 ns. This
simulation length is two orders of magnitude larger than the lengths used in the seminal
work of Espinosa and coworkers to distinguish if
trajectories show or not induction periods. Why? Because there
exist subtle but key differences between formation of Ih ice from
liquid water at coexistence conditions and the formation of \ch{CO2} hydrates from \ch{CO2} dissolved in water, also at coexistence conditions. In the case of the original implementation of the MI methodology, water molecules can accommodate nearly
immediately in the crystallographic positions of the Ih ice structure when the mold is switched on. In the current case, the stable crystalline phase is only formed when the guest molecules (\ch{CO2}), occupy their equilibrium positions and water molecules are rearranged around the \ch{CO2} molecules to enclathrate them in the corresponding hydrate cages. Therefore, formation of \ch{CO2} hydrates is dominated by thermodynamic and kinetic hindrances.
The \ch{H2O} + \ch{CO2} binary mixture exhibits LL immiscibility at thermodynamic conditions at which \ch{CO2} hydrate is formed. The low solubility of \ch{CO2} in water acts as thermodynamic hindrance, increasing simulation times. In addition to that, diffusion of \ch{CO2} (from the \ch{CO2}-rich liquid phase) across the \ch{H2O}-rich liquid phase acts as kinetic hindrance, that also increases simulation times. As a consequence of this, longer simulation runs are needed to distinguish different behaviors of $n_{h}$, as a function of time, and to identify correctly possible induction periods in trajectories.

It is possible to identify three different scenarios that allow to
determine with confidence the optimal value of $r_{w}$, its uncertainty, and consequently, $\gamma_{hw}$ and its error: scenario I, corresponding to
low values of $r_{w}$, in which none of the trajectories exhibits
induction period and system always crystallizes if simulation runs
are sufficiently long (no free energy barrier between the fluid and solid phases exists); scenario II, corresponding to intermediate values of $r_{w}$, in which at least one of the independent trajectories exhibits an induction period, and after some time, the system crystallizes. In addition to that, all independent trajectories finally crystallize if the simulation runs are sufficiently long. The particular length that a given simulation must run depends on the system. Although in this case there is an energy barrier between the fluid and the solid slab, the system is able to overcome it at some point; and finally, scenario III, corresponds to large values of $r_{w}$. In this case, none of the trajectories shows, on average, monotonous increase of $n_{h}$ with time. This indicates that there is
a free energy barrier between the fluid and the solid high enough
that does not allow the system to crystallize.

As can be seen in Figure~\ref{nh_vs_t} (see SI material for the rest of $r_{w}$ investigated), simulations with $r_{w} \leq 1.172\,\text{\AA}$ (0.37$\sigma$) do not show
any induction period for all the trajectories considered since $n_{h}$ grows, on average, monotonically from $t \approx 0$ ns. In all cases, there is not free energy barrier between the liquid water and the solid \ch{CO2} hydrate slab. Consequently, these simulations cannot be used for thermodynamic integration because the integration path crosses a first-order phase transition (scenario I). For simulations with $1.203\,\text{\AA} \leq r_{w} \leq 1.283\,\text{\AA}$ ($0.38\sigma \leq r_{w} \leq 0.405\sigma$), at least one trajectory shows an induction period, i.e., $n_{h}$ does not grow, on average, from $t \approx 0$ ns. Particularly, different induction periods can be observed for systems with $r_{w} = 1.203$ (0 -- 30 ns
for orange and green trajectories), 1.219 ( 0 -- 30 ns for blue trajectory and 0 -- 70 ns for orange trajectory), 1.235 (0 -- 10 ns for orange, 0 -- 15 ns for maroon, and 0 -- 20 ns for blue
trajectories), 1.267 (0 -- 40 ns for red, 0 -- 55 ns for green, and 0 -- 65 ns for orange trajectories), and 1.283$\,\text{\AA}$ (0 -- 45 ns for green trajectory). The common behavior of these systems,
according to scenario II, is that all trajectories exhibit, on average, a monotonous growing of $n_{h}$ with time, indicating that although systems have free energy barriers between the liquid and solid phases, they are able to overcome them and crystallize if simulations are long enough. Finally, none of the simulations with $r_{w} \geq 1.298\,\text{\AA}$ ( $r_{w} \geq 0.41\sigma$) exhibit, on average, monotonic increasing of $n_{h}$ with time due to the existence of a free energy barrier between the fluid and solid that does not allow the system to crystallize.

According to this and following Espinosa \textit{et al.}, $r^{0}_{w}$ should be between $r_{w}= 1.172\,\text{\AA}$ and $r_{w} = 1.203 \,\text{\AA}$. However, the determination of interfacial free energies of hydrates at coexistence suffers thermodynamic and kinetic hindrances that makes difficult to distinguish the behavior of the system, as it has already been explained. Due to this and to ensure that $r^{0}_{w}$ and consequently $\gamma_{hw}$, are provided with appropriate uncertainties, we have chosen a more conservative confidence interval. We assume that $r^{(l)}_{w}= 1.140\,\text{\AA}$  is the lower bound of $r^{0}_{w}$ and $r^{(u)}_{w}= 1.235\,\text{\AA}$ its upper bound according to the mold integration method. With this election, $r^{0}_{w} = \frac{r^{(l)}_{w}+r^{(u)}_{w}}{2} \approx 1.1875 \,\text{\AA}$ the mean value of the upper and lower bounds of $r_{w}$. The uncertainty of $r^{0}_{w}$ is estimated as half of the distance between the $r^{(u)}_{w}$ and $r^{(l)}_{w}$, i.e.,
$\sigma_{r^{0}_{w}} = \frac{r^{(u)}_{w}-r^{(l)}_{w}}{2} \approx 0.0475\,\text{\AA}$ . Keeping only the
most significant figure for the uncertainty and rounding off
appropriately, the final value of the optimal radius is $r^{0}_{w} =1.19(5) \,\text{\AA}$.

\textbf{Thermodynamic integration.} Once $r_{w}^{0}$ is known, we determine $\gamma_{hw}(r_{w})$ by means of thermodynamic integration for different values of $r_{w}>r_{w}^{0}$. According to the results presented in the previous section, $r_{w}=1.267$, $1.283$, and $1.298\,\text{\AA}$ are appropriate values for the determination of the \ch{CO2} hydrate -- water interfacial free energy using the MI methodology.

To calculate the integral of Eq.~(\ref{deltaG_m}) for each potential range of interacting sites of the model, $r_{w}$, it is necessary to evaluate the average of filled wells with water
molecules, $\langle N_{fw}(\varepsilon) \rangle_{_{NP_{z}\mathcal{A}T}}$, at a given number of $\varepsilon$ values. Due to the nature of the interaction potential between the sites of the mold and water molecules this is simply the total mold-water interaction energy divided by $-\varepsilon$. We have checked that accurate values of the integral are obtained using typically 15 different values of $\varepsilon$. Each point has been obtained switching the mold on with the corresponding well-water interaction parameter value $\varepsilon$ and performing MD simulations
in the $NP_{z}\mathcal{A}T$ ensemble. Simulations are equilibrated during $20\,\text{ns}$ and $N_{fw}(\varepsilon)$, the number of filled wells, is averaged over $50\,\text{ns}$. Note that simulation periods used in this work are longer than those used in previous works of Espinosa \textit{et al.}~\cite{Espinosa2016a, Espinosa2014a}. This is because, as explained in the previous section, \ch{CO2} molecules must diffuse along the whole simulation box to occupy their equilibrium positions in order to form a stable hydrate slab in the middle of the box. The number of filled wells, $N_{fw}$, are obtained as appropriate averages during the production period. In order to estimate errors $N_{fw}$, we have applied the sub-block method. The production
period, 50 ns, is divided into $M$ independent blocks. The statistical errors are then estimated from the standard deviation of the average, $\sigma_{<N_{fw}>} = \overline{\sigma_{fw}}/\sqrt{M} $ where $\overline{\sigma_{fw}}$ is the variance of the block average and M has been fixed to $M = 10$.

\begin{figure}
\centering
\includegraphics[angle=0,width=1.10\columnwidth]{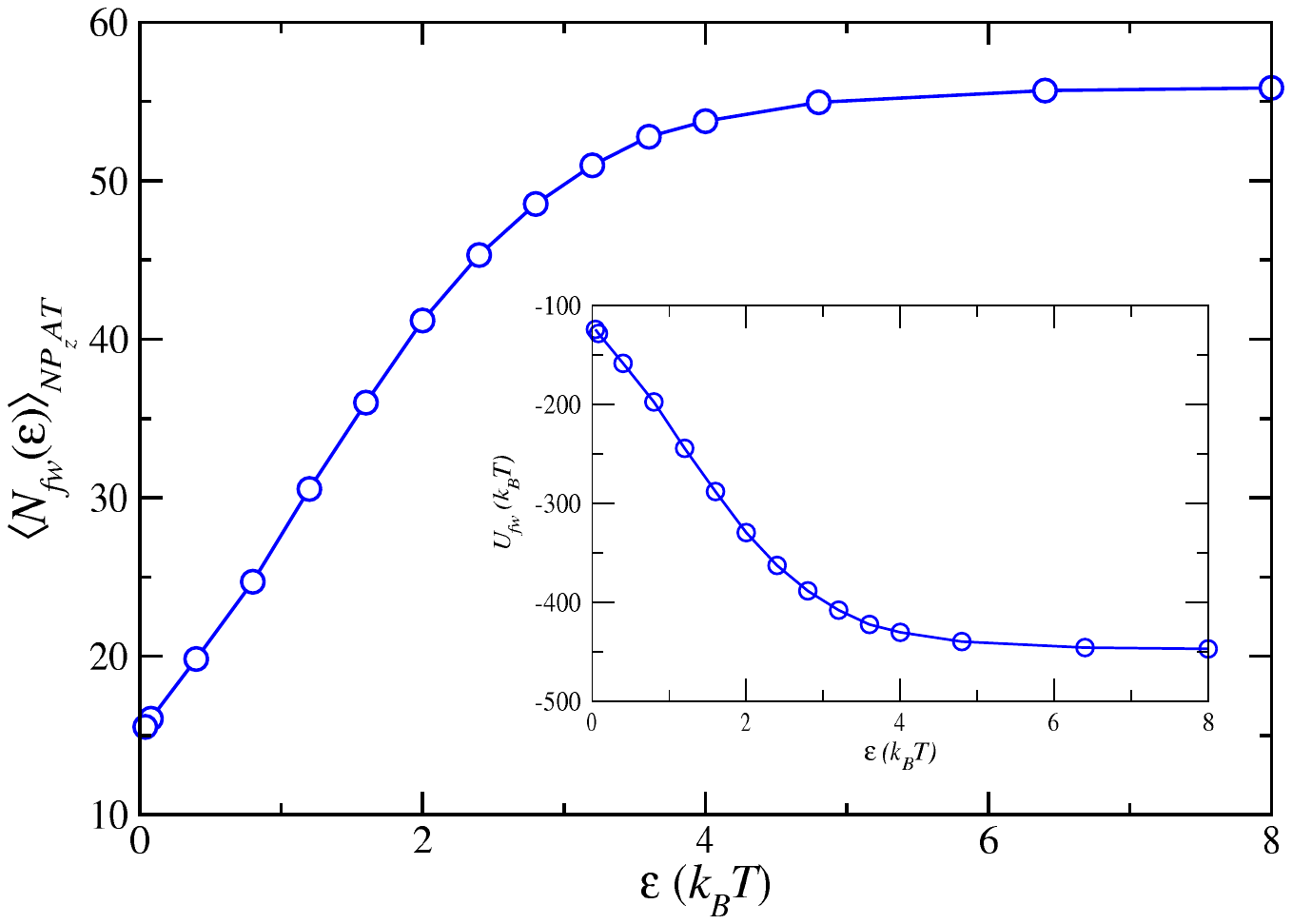}
\caption{\small Averaged number of filled wells, $\langle N_{fw}(\varepsilon) \rangle_{_{NP_{z}\mathcal{A}T}}$, as a function of the well depth $\varepsilon$, for the principal plane of the \ch{CO2} hydrate. The radius of the mold used is $1.283\,\text{\AA}$. All simulations are performed at $40\,\text{MPa}$ and $287\,\text{K}$. The circles correspond to the value obtained from $NP_{z}\mathcal{A}T$ simulations with $20\,\text{ns}$ of equilibration and $80\,\text{ns}$ of production. The inset represents the $U_{fw}(k_{B}T)$ potential energy as a function of the well depth $\varepsilon$. Note that $N_{fw}=U_{fw}/(-\varepsilon)$. The curves are included as a guide to the eye.}
\label{N_vs_epsilon}
\end{figure}

As an example, Figure~\ref{N_vs_epsilon} shows the $\langle N_{fw}(\varepsilon) \rangle_{_{NP_{z}\mathcal{A}T}}$ behavior as a function of the well depth, $\varepsilon$, for the case $r_{w}=1.283\,\text{\AA}$. Results corresponding to the other interaction ranges ($1.267$ and $1.298\,\text{\AA}$) exhibit the same qualitative behavior that those presented here.  
The error bars cannot be distinguished in Figure~\ref{N_vs_epsilon}  because are
smaller than the symbol size. Particularly, $\sigma_{<N_{fw}>} =0.04$ for low values of the well depth, $\varepsilon \approx k_{B}T$ and $\sigma_{<N_{fw}>} =0.03$ for high
values of the well depth, $\varepsilon \approx 6 - 7 \,k_{B}T$.
$\langle N_{fw}(\varepsilon) \rangle_{_{NP_{z}\mathcal{A}T}}$ increases smoothly as the strength of the well-water interacting potential is increased. For $\varepsilon=0$, when the mold is switched off, the average number of wells occupied is $16$, approximately. As $\varepsilon\rightarrow\varepsilon_{m}$ (mold fully switched on), the average number of wells occupied tends to $56$, the total number of well available in the simulation box, $N_{W}$. Note that once the plateau is reached,  $\langle N_{fw}(\varepsilon) \rangle_{_{NP_{z}\mathcal{A}T}}$, and consequently $\gamma_{hw}(r_{w})$, is independent of $\varepsilon$. It is interesting to remark that the maximum $\varepsilon$ value used in this work, $\varepsilon=8k_{B}T$, satisfies the two main conditions that ensure the goodness of the MI methodology: (1) every well of the mold is occupied by a water molecule in the final state of the thermodynamic integration, i.e., $\langle N_{fw}(\varepsilon) \rangle_{_{NP_{z}\mathcal{A}T}}\,\rightarrow N_{w}$ as $\varepsilon\rightarrow\varepsilon_{m}$; and
(2) the $\varepsilon_{m}$ value chosen in this work guarantees the smoothness of the function $\langle N_{fw}(\varepsilon) \rangle_{_{NP_{z}\mathcal{A}T}}$.
Particularly, we have checked that other values, such as $\varepsilon = 6$ and even $16 \, k_{B}T$, provide identical results for interfacial energy values, providing essentially the same curve as that shown in Figure~\ref{N_vs_epsilon}.

\begin{figure}
\centering
\includegraphics[angle=0,width=1.10\columnwidth]{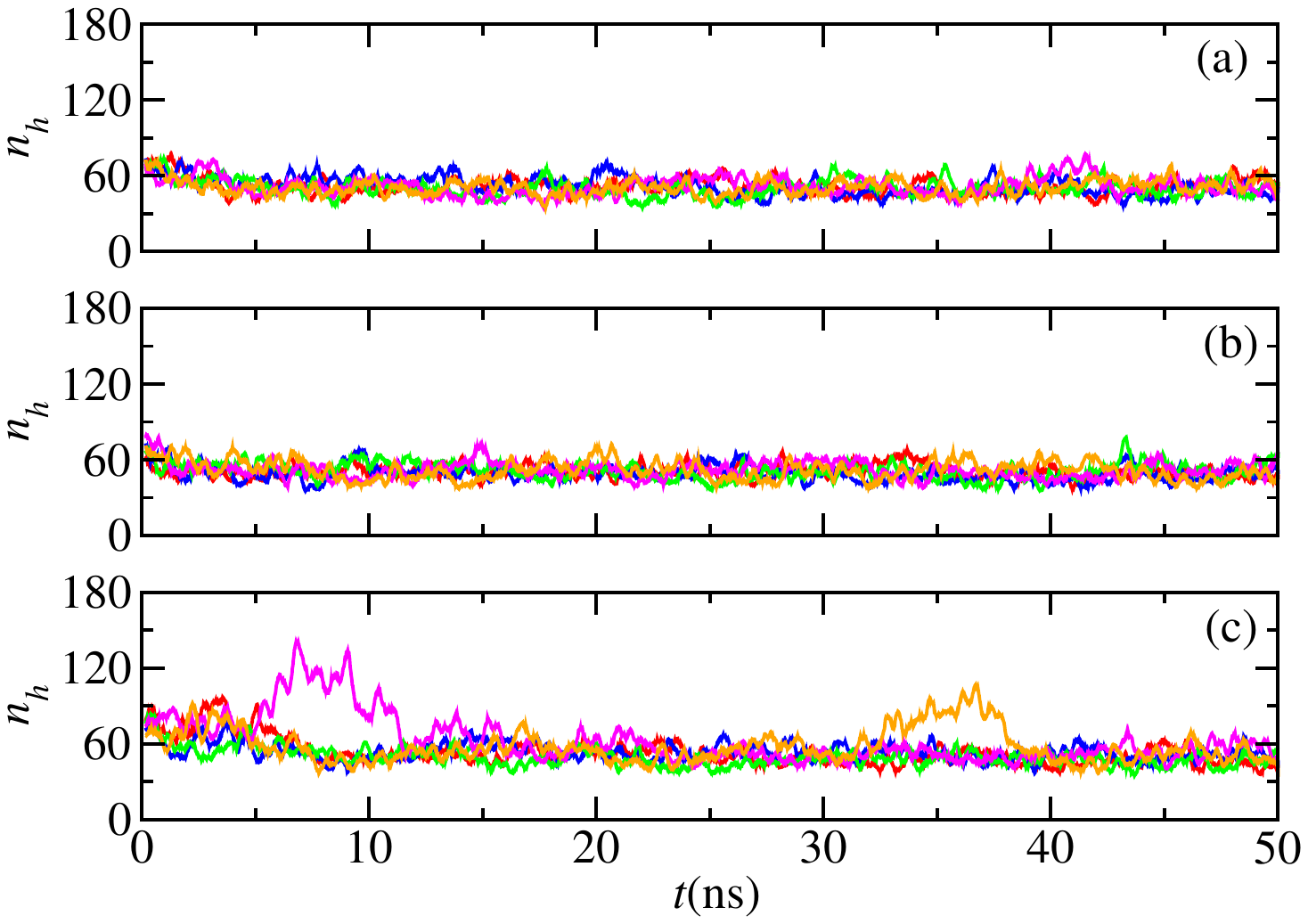}
\caption{\small Number of water molecules in the crystal slab, $n_{h}$, as a function of time for the principal plane of the \ch{CO2} hydrate. Different curves correspond to the well depths analyzed in Figure~\ref{N_vs_epsilon}: (a) $0.04$ (red), $0.08$ (blue), $0.4$ (green), $0.8$ (magenta), and $1.2$ (orange); (b) $1.6$ (red), $2$ (blue), $2.4$ (green), $2.8$ (magenta), and $3.2$ (orange); and (c) $3.6$ (red), $4$ (blue), $4.8$ (green), $6.4$ (magenta), and $8\,k_{B}T$ (orange). The radius of the mold used is $r_{w}=1.298\,\text{\AA}$. All the results have been obtained using $NP_{z}T\mathcal{A}$ simulations at $40\,\text{MPa}$ and $287\,\text{K}$.}
\label{N_vs_t}
\end{figure}

As we have previously mentioned, all thermodynamic states visited by the system during the simulations must be in the liquid phase. This ensures that the integration path to determine $\Delta G_{m}$ does not cross a first-order phase transition, and consequently, $\Delta G_{m}$ can be easily related with $\Delta G^{hw}$, the reversible work to form the crystalline slab. This can be done monitoring the number of water molecules that belong to the hydrate slab, $n_{h}$, and checking that it does not grow indefinitely at any integration point. Figure \ref{N_vs_t} shows the evolution of $n_{h}$ as a function of time for
all the $\varepsilon$ values analyzed for $r_{w}=1.298\,\text{\AA}$. As can be seen, in nearly all cases $n_{h}$ show a stationary value around 50--60 water molecules in the simulation box. Particularly,
$n_{h}$ does not grow beyond $90$ water molecules, approximately. Note that we are using $56$ interacting sites for the mold and the total number of water molecules used in simulations is $736$, i.e., the hydrate phase does not grow irreversibly for any of the well depths and associating energies used during the thermodynamic integration ensuring the reversibility of the simulations. This indicates that there is a stable hydrate slab formed in the middle of the simulation box and two \ch{CO2} hydrate -- water interfaces. It is interesting to note that there is an incipient and maintained crystal growing for the deepest wells, $\varepsilon=6.4$ and $8k_{B}T$ corresponding to the magenta
and orange curves in part (c) of Figure~\ref{N_vs_t}, respectively. However, these fluctuations are only maintained during $5\,\text{ns}$ in both cases, approximately. After this time, the average number of water molecules in the hydrate crystalline phase fluctuates around 50--60 molecules. This indicates clearly that simulations for all the wells, including those for $6.4$ and $8\,k_{B}T$, are performed at conditions at which thermodynamic integration can be safely performed.

\textbf{Interfacial free energy.} From the behavior of the averaged number of filled wells, $\langle N_{fw}(\varepsilon) \rangle_{_{NP_{z}\mathcal{A}T}}$, as a function of the well depth, $\varepsilon$, it is possible to determine the interfacial free energy of the \ch{CO2} hydrate -- water interface for a given value of the
radius of the mold used to perform our calculations. In particular, for the case of $r_{w}=1.267\,\text{\AA}$, we have determined a value
of $\Delta G_{m}=-372.51\,k_{B}T$. According to Eq.~\eqref{gamma_hw2}, one must subtract the well -- oxygen associating energy to obtain the reversible work needed to form the hydrate crystalline slab, $\Delta G^{hw}=\Delta G_{m} + N_{w}\varepsilon_{m}=
-372.51\,k_{B}T+56\times 8\,k_{B}T=75.49\,k_{B}T$. Using the definition of the interfacial free energy given by Eq.~(\ref{gamma_hw1}), the \ch{CO2} hydrate interfacial energy obtained from the MI technique with $r_{w}=1.267\,\text{\AA}$ is $\gamma_{hw}=25.89\,\text{mJ/m}^{2}$. The same direct calculation can be done for the rest $r_{w}$ values employed in this work. We obtain $\gamma_{hw}=25.52\,\text{mJ/m}^{2}$ for $r_{w}=1.283\,\text{\AA}$ and $\gamma_{hw}=24.77\,\text{mJ/m}^{2}$ for $r_{w}=1.298\,\text{\AA}$.

\begin{figure}[h!]
\centering
\includegraphics[angle=0,width=1.10\columnwidth]{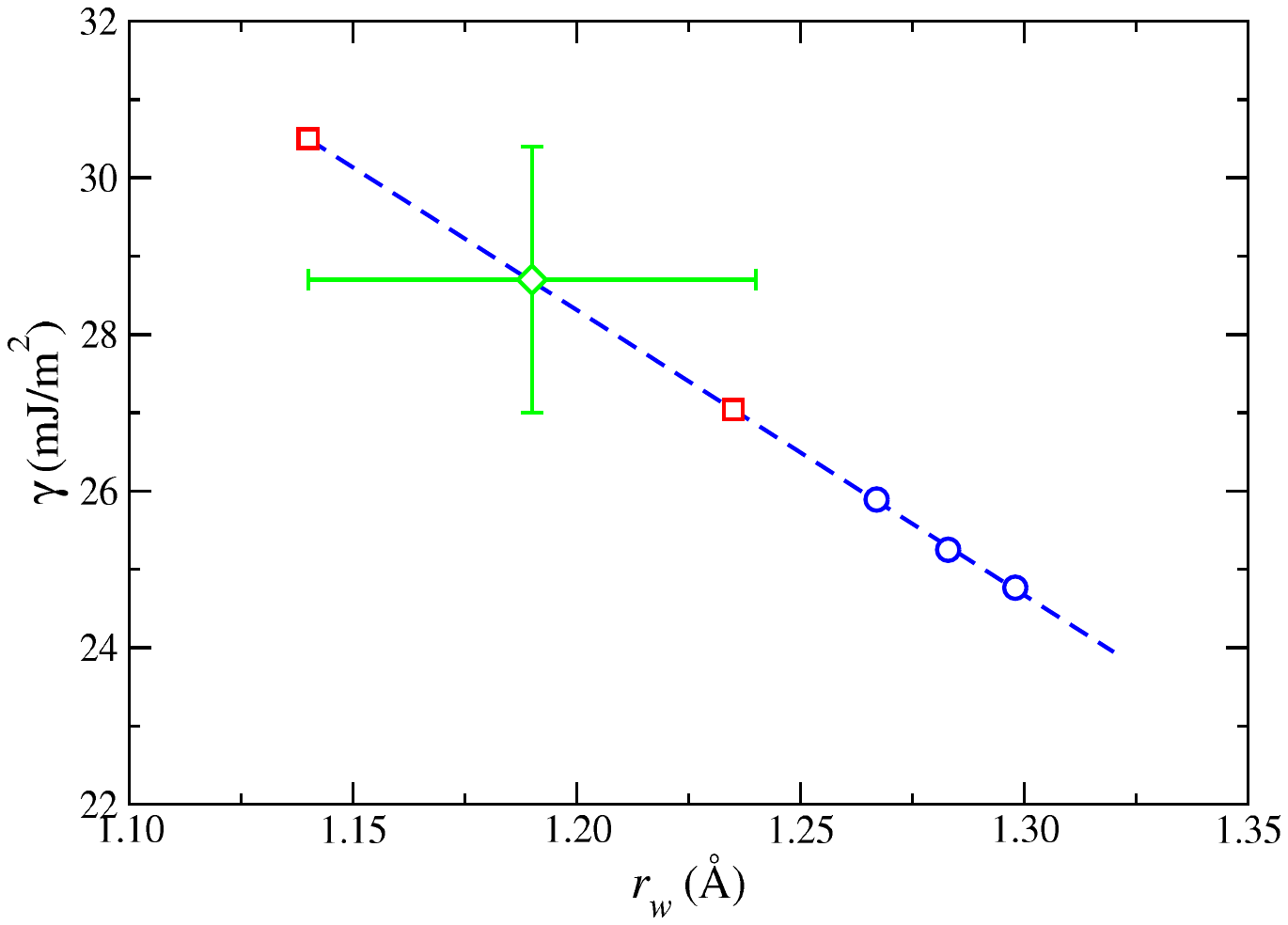}
\caption{\small Interfacial free energy, as a function of the potential well radius of the mold (open blue circles). The blue dashed line represents a linear fit of the data, the open red squares the interfacial tension evaluated at $r^{(l)}_{w}$ and $r^{(u)}_{w}$, and the open green diamond the extrapolation of the linear fit to the optimal well radius.}
\label{gamma_vs_rw}
\end{figure}

Following the works of Espinosa \textit{et al.}~\cite{Espinosa2016a,Espinosa2014a, Espinosa2015a}, we have represented the $\gamma_{hw}$ values obtained for three different values of $r_{w}>r_{w}^{0}$. Figure~\ref{gamma_vs_rw} shows the values of $\gamma_{hw}$ as a function of $r_{w}$. As can be seen, data follow a linear dependence of the interfacial free energy with $r_{w}$, as in previous works. Although there is not a definite proof for this behavior, our results confirm that the interfacial free energy of the \ch{CO2} hydrate-water interface follows the same trend (dashed blue line).
According to the approach presented by Espinosa and coworkers, we perform a linear regression using the three $\gamma_{hw}$
values obtained in this work and determine the interfacial free energy of the system, i.e., $\gamma_{hw}$ evaluated at $r_{w}=r_{w}^{0}=1.19\,(5)\,\text{\AA}$. From the results presented in Figure~\ref{gamma_vs_rw},
the interfacial free energy is obtained as the mean value
of $\gamma_{hw}$ evaluated at the lower bound of $r_{w}$, $r^{(l)}_{w}= 1.140\,\text{\AA} $ ($\gamma^{(l)}_{hw} =30.50 \,\text{mJ/m}^{2}$) and the upper bound of $r^{(u)}_{w}= 1.235\,\text{\AA}$  ($\gamma^{(u)}_{hw} =27.04 \,\text{mJ/m}^{2}$), given a value $\gamma_{hw}=(\gamma^{(l)}_{hw} + \gamma^{(u)}_{hw})/2 = 28.67 \,\text{mJ/m}^{2}$ (open red squares). Uncertainties associated to $\gamma_{hw}$ come from
two sources: errors due to the estimations of $r_{w}^{0}$ and uncertainties associated with the calculation of each $\langle N_{fw}(\varepsilon) \rangle$ in the integral
of Eq.~\eqref{deltaG_m}. Unfortunately, it is difficult to estimate the propagation of uncertainties to the integral defined by $\Delta G_{m}$. Hence, we assume that the main error in $\gamma_{hw}$ is due to the uncertainty in $r_{w}^{0}$ . This
uncertainty is estimated as follows. Since the value of $\gamma_{hw}$ is in the middle of the interval $[\gamma^{(l)}_{hw}, \gamma^{(u)}_{hw}]$, we assume the uncertainty of $\gamma_{hw}$ is $\sigma_{\gamma_{hw}} =(\gamma^{(u)}_{hw}-\gamma^{(l)}_{hw})/2 \approx 1.73 \,\text{mJ/m}^{2}$. Considering only the most significant figure for the uncertainty and rounding off appropriately, the uncertainty of $\gamma_{hw}$ is $\sigma_{\gamma_{hw}}= 2 \,\text{mJ/m}^{2}$. With this
election, the final value of the interfacial free energy and its
uncertainty is $\gamma_{hw}= 29(2) \,\text{mJ/m}^{2}$ (open green diamond).

It is possible to compare the \ch{CO2} hydrate interfacial free energy obtained by means of computer simulation with experimental data taken from the literature. According to the best of our knowledge, Uchida \textit{et al.} reported an interfacial free energy value $\gamma_{hw}=28\,(6)\,\text{mJ/m}^{2}$ for the \ch{CO2} hydrate-water interface (after the correction of Anderson \textit{et al.} explained previously). Anderson and collaborators obtained a slightly different value, $\gamma_{hw}=30\,(3)\,\text{mJ/m}^{2}$ for the same system.

The origin of the discrepancies between the Uchida and Anderson experiments can be found in the use of different values of the magnitudes involved in the Gibbs--Thomson equation~\cite{Kowalsky2007a,Tsimpanogiannis2019a}. Additionally, its use assumes that all pores in the porous materials are cylindrical, with only one and well-defined pore size, independent, and non-connected~\cite{Clennell1999a,Henry1999a}. Unfortunately, real mesoporous materials are far from this idyllic picture. They are made by a network of interconnected cylindrical pores with a given pore size distribution~\cite{Gelb1999a}. This produces a combination of ``individual'' experimental dissociation curves and different definitions for $d$, the mean pore diameter. Finally, there exists an unfrozen layer of liquid (water), of few angstroms in thickness, between the solid phase (hydrate) and the pore wall~\cite{Uchida1999a,Uchida2002a,Anderson2003a,Anderson2003b,Churaev1993a,Ishikiriyama1995a,Faivre1999a,Radhakrishnan2002a}, producing an enhancement of the interfacial free energies that is is difficult to quantify and correct, and consequently the values are slightly larger than those in absence of pores.

The results obtained in this work are relevant due to several reasons. First at all, the value obtained for the hydrate-water interfacial free energy here, $\gamma_{hw}=29(2)\,\text{mJ/m}^{2}$, is within the experimental value measured by Uchida \textit{et al.}~\cite{Uchida2002a},
$\gamma_{hw}=28(6)\,\text{mJ/m}^{2}$. The experimental determination of Anderson \textit{et al.} for the interfacial energy ranges from $27$ to $33\,\text{mJ/m}^{2}$. Our value is also
within the boundaries provided by these authors. Note the values for $\gamma_{hw}$ obtained are reliable since the same model for water (TIP4P/ice) has been recently used by Vega and coworkers to accurately determine the Ih ice-water interfacial free energy~\cite{Espinosa2016a}. Taking into account the absence of more reliable experimental values (estimated error of Uchida's value are around $\sim 20\%$ and those of Anderson $\sim 10\%$), and the difficulties for determining experimentally and using computational techniques SF interfacial free energies, we can conclude that agreement
between computer simulation predictions and experimental data is excellent.

Not only agreement between predictions from computer simulation and experimental data taken from the literature is outstanding. From a more fundamental point of view, the estimation of the hydrate-water interfacial energy presented in this paper is only based on basic principles without the use of phenomenological theories. Particularly, the predictions are obtained from: (1) the thermodynamic definition of the interfacial free energy, i.e., as the reversible work needed to create a hydrate-water interface at coexistence; (2) the solution of the Newton's classical equations of motion of the molecules that form the system using well-established MD techniques; and (3) the use of rotationally invariant local bond-order parameters that allow to
distinguish between hydrate-like and water--like molecules during computer simulations.

This approach contrasts with other theoretical methods or even experiments themself that use phenomenological theories or approximations. For instance, the Seeding technique proposed by Bai and Li~\cite{Bai2006a} has been recently used to determine the interfacial energy of different molecular models, including water and hydrates\cite{Knott2012a,Sanz2013a}. Care must be taken with this methodology since the interfacial energy value does not correspond to the that at coexistence conditions since the technique only works at supersaturation conditions, i.e., at temperatures well below the coexistence temperature. To obtain the interfacial energy value at coexistence it is necessary to extrapolate the values obtained to the coexistence temperature. This approach uses the Classical Nucleation Theory to estimate the interfacial free energy assuming important
approximations, including that the critical cluster obtained during simulation is spherical. In the case of the Uchida \textit{et al.} and Anderson \textit{et al.} experiments, interfacial values are indirect measurements with a number of approximations that could provoke significant uncertainties in the measurements, as it has been previously explained: all pores in silica glass are cylindrical, existence of a finite distribution of pore sizes, and use of different values for the enthalpy for hydrate dissociation and hydrate density, among other assumptions.

It is relevant to comment in this context that
Phan and collaborators~\cite{Phan2022a} have recently
determined the cyclopentane hydrate-water interfacial energy from experiments and
compared with computer simulations that determine the contact angle of a water droplet on a cyclopentane immersed in cyclopentane. This work is the continuation of a previous work~\cite{Phan2021a} in which authors study via experiments the wettability and hydrate film growth of a cyclopentane hydrate in pure cyclopentane. Particularly interesting are the results corresponding to predictions from simulation since show an alternative way to determine interfacial free energies of hydrate-water interfaces. This is relevant here since the cyclopentane hydrate-water interfacial energy found by computer simulation, $\gamma_{hw}=0.24(16)\,\text{mJ/m}^{2}$, agrees very well with experiments performed by the same authors, $0.32(5)\,\text{mJ/m}^{2}$. It is
also interesting to check that this value is two orders of magnitude lower that than of the CO$_{2}$ hydrate-water interfacial energy found in this work, $29(2)\,\text{mJ/m}^{2}$.

Finally, it is important to recall here one of the drawbacks of using porous materials to measure interfacial free energies of solids in combination with the Gibbs--Thomson relationship. As it was mentioned, it is well-known that experiments of dissociation of hydrates in porous materials involve the presence of an unfrozen layer of water, of few angstroms in thickness, between the hydrate and the
pore wall~\cite{Uchida1999a,Uchida2002a,Anderson2003a,Anderson2003b,Churaev1993a,Ishikiriyama1995a,Faivre1999a,Radhakrishnan2002a}. This provokes an enhancement of the interfacial free energies, and consequently the values measured are slightly larger that those in the bulk phase due to the presence of the pore walls. The overestimation of the interfacial free energy value obtained from this technique, with respect to that in bulk, could be around $1-5\,\text{mJ/m}^{2}$,
approximately~\cite{Schreiber2001a,Jahnert2008a,Kittaka2011a,Morishige2018a,Yao2019a}. However, our simulation results seem to indicate that this effect is not as high in the case of the ice-water interface. Note that in this work there is no unfrozen layer of liquid water since we are calculating the interfacial free energy in the bulk.

\section{Conclusions}

We have determined the \ch{CO2} hydrate -- water interfacial free energy by means of computer simulation using reliable molecular models for water and carbon dioxide. This is the first calculation of the \ch{CO2} hydrate interfacial energy using molecular dynamics simulations at coexistence conditions from fundamental principles, including the definition of interfacial free energy, Thermodynamics, and Statistical Mechanics.

To predict the interfacial free energy, we use the recently proposed mold integration method~\cite{Espinosa2014a,Espinosa2016a}  to deal with \ch{CO2} hydrates. In order to apply the methodology correctly, we explicitly take into account that: (1) we are dealing with binary mixtures; (2) we need to explicitly simulate the coexistence of three phases, hydrate, aqueous solution of water and aqueous solution of carbon dioxide; and (3) rotationally invariant local bond order parameters~\cite{Lechner2008a} have to be optimized to identify water-like and hydrate-like particles and to follow the growth of the hydrate solid phase along the simulations.

We hypothesize that accurate models for water and CO$_{2}$, in combination with advanced simulation techniques, are able to account for good estimates of the CO$_{2}$ hydrate-water interfacial energy. Particularly, we use the well-know TIP4P/ice model for water~\cite{Abascal2005b} and the TraPPE forcefield for CO$_{2}$~\cite{Potoff2001a}. We think the TIP4P/ice model is a good choice since is able to correctly predict the ice-water interfacial energy. According to Conde and Vega~\cite{Conde2010b,Conde2013a}, water models that accurately predict the melting point of ice are good candidates for predicting the dissociation line of hydrates. In addition to that, the combination of the TIP4P/ice and the TraPPE CO$_{2}$ models is also a good hypothesis since they are able to account correctly for the dissociation line of the hydrate~\cite{Miguez2015a}, a plausible condition if its interfacial energy has be to estimated at coexistence. 

The interfacial free energy predictions obtained in this work are compared with the two only indirect experimental data existing in the literature from Uchida \textit{et al.}~\cite{Uchida1999a, Uchida2002a} and Anderson \textit{et al.}~\cite{Anderson2003a, Anderson2003b} that combined hydrate equilibrium measurements under confinement through the Gibbs-Thomson equation~\cite{Handa1992a,Clennell1999a,Henry1999a}. According to the authors, both results are consistent but they should be considered as conservative estimates since depend on available physical property data, including enthalpy for hydrate dissociation and hydrate density, and the use of a phenomenological equation. The interfacial free energy value obtained from simulation, $29(2)\,\text{mJ/m}^{2}$, agrees equally well with the experimental results of Uchida \emph{et al.}, $28(6)\,\text{mJ/m}^{2}$, and Anderson \emph{et al.}, $30(3)\,\text{mJ/m}^{2}$. The value of obtained from simulation also suggests that the possible enhancement of the interfacial free energies due to the presence of an unfrozen layer of water between hydrates and pore walls could be not as high as in
the case of the ice-water interface~\cite{Schreiber2001a,Jahnert2008a,Kittaka2011a,Morishige2018a,Yao2019a}.

Computer simulation can be considered a valuable complementary method to provide reliable and confident interfacial free energy values for hydrates. Our calculations show that the mold integration methodology can be used to determine interfacial free energies of complex solid phases, including hydrates. Our pioneering work demonstrates that is possible to predict interfacial energies of hydrates
from a truly computational molecular perspective and opens a new door to the determination of free energies of hydrates.

\section*{CRediT authorship contribution statement}

%%Conceptualization, Methodology, Software, Validation, Formal analysis, Investigation, Data curation, Writing – original draft, Visualization, Resources, Supervision, Project administration, Funding acquisition, Writing – review $\&$ editing.

\textbf{Jes\'us Algaba:} Methodology, Investigation.
\textbf{Esteban Acu\~na:} Validation, Investigation.
\textbf{Jos\'e Manuel M\'iguez:} Conceptualization, Supervision.
\textbf{Bruno Mendiboure:} Investigation, Formal analysis.
\textbf{Iv\'an M. Zer\'on:} Writing – review $\&$ editing.
\textbf{Felipe J. Blas:} Funding acquisition, Writing – review $\&$ editing.

\section*{Declaration of Competing Interest}
The authors declare that they have no known competing financial interests or personal relationships that could have appeared  to influence the work reported in this paper.

\section*{Acknowledgements}

We thank Pedro J. P\'erez for the critical reading of the manuscript.
We also acknowledge Centro de Supercomputaci\'on de Galicia
(CESGA, Santiago de Compostela, Spain) and MCIA
(M\'esocentre de Calcul Intensif Aquitain) of the Universit\'es de
Bordeaux and Pau et Pays de l'Adour (France) for providing
access to computing facilities. We thank financial support from the Ministerio de Econom\'ia, Industria y Competitividad (FIS2017-
89361-C3-1-P), Junta de Andaluc\'ia (P20-00363), and
Universidad de Huelva (P.O. FEDER UHU-1255522), all three cofinanced by EU FEDER funds. J.A. acknowledges Contrato
Predoctoral de Investigaci\'on from XIX Plan Propio de
Investigaci\'on de la Universidad de Huelva and a FPU Grant (Ref.
FPU15/03754) from Ministerio de Educaci\'on, Cultura y Deporte.
J. A., J. M. M., and F. J. B. thankfully acknowledge the computer
resources at Magerit and the technical support provided by the
Spanish Supercomputing Network (RES) (Project QCM-2018-2-
0042).

%% The Appendices part is started with the command \appendix;
%% appendix sections are then done as normal sections

\appendix
\section*{Supporting Information}
%\label{sec:sample:appendix}

See Supporting Information for a general perspective of the $n_{h}$ time evolution for complete set of $r_{w}$ investigated.

%% If you have bibdatabase file and want bibtex to generate the
%% bibitems, please use
%%
 \bibliographystyle{elsarticle-num} 
 \bibliography{bibfjblas}

@string{aiche = {AIChE Journal.}}

@string{ae = {Appl. Energy}}

@string{cej = {Chem. Eng. J.}}

@string{cg = {Chem. Geol.}}

@string{cgd = {Cryst. Growth Des.}}

@string{csa = {Colloids Surf. A}}

@string{ecm = {Energy Conversion and Management}}

@string{epjb = {Eur. Phys. J. B}}

@string{fpe = {Fluid Phase Equil.}}

@string{fuel = {Fuel}}

@string{ijggc = {Int. J. Greenhouse Gas Control}}

@string{jacs = {J. Am. Chem. Soc.}}

@string{jced = {J. Chem. Eng. Data}}

@string{jcej = {J. Chem. Eng. Jpn.}}

@string{jcp = {J. Chem. Phys.}}

@string{jctc = {J. Chem. Theory Comput.}}

@string{jcis = {J. Colloid Interf. Sci.}}

@string{jgr = {J. Geophys. Res.}}

@string{jngse = {J. Nat. Gas Sci. Eng.}}

@string{jpc = {J. Phys. Chem.}}

@string{jpcb = {J. Phys. Chem. B}}

@string{jpcc = {J. Phys. Chem. C}}

@string{la = {Langmuir}}

@string{molphys = {Mol. Phys.}}

@string{pccp = {Phys. Chem. Chem. Phys.}}

@string{prb = {Phys. Rev. B}}

@string{prevlet = {Phys. Rev. Lett.}}

@string{rpp = {Rep. Prog. Phys.}}

@string{science = {Science}}

@article{Miguez2015a,
Author = {J. M. M{\'\i}guez and M. M. Conde and J.-P. Torr{\'e} and F. J. Blas and M. M. Pi{\~n}eiro and C. Vega},
title={Molecular Dynamics Simulation of \ch{CO2} Hydrates: Prediction of Three Phase Coexistence Line},
Journal = jcp,
Pages = {124505-1--124505-12},
Volume = {142},
doi = {https://doi.org/10.1063/1.4916119},
Year = {2015}}

@article{Gelb1999a,
	Author = {L. D. Gelb and K. E. Gubbins and R. Radhakrishnan and M. Sliwinska-Bartkowiak},
Title = {Phase Separation in Confined Systems},
	Journal = rpp,
	Pages = {1573},
	Volume = {62},
	doi = {https://doi.org/10.1088/0034-4885/62/12/201},
	Year = {1999}}

@article{Essmann1995a,
Author = {U. Essmann and L. Perera and M. L. Berkowitz and T. Darden and H. Lee and L. G. Pedersen},
Title = {A Smooth Particle Mesh Ewald method},
Journal = jcp,
Pages = {8577-8593},
Volume = {103},
Year = {1995},
doi = {https://doi.org/10.1063/1.470117}}

@book{Frenkel2002a,
	Author = {D. Frenkel and B. Smit},
	Publisher = {2nd Ed. Academic, San Diego},
	Title = {Understanding Molecular Simulations},
	Year = {2002}}

@book{Allen2017a,
Author = {M. P. Allen and D. J. Tildesley},
Publisher = {Oxford University PressClaredon, Oxford},
Title = {Computer Simulation of Liquids, 2 Ed.},
Year = {2017}}

@article{Nose1984a,
Author = {S. Nos{\'e}},
Title={A Molecular Dynamics Method for Simulations in the Canonical Ensemble},
Journal = molphys,
Pages = {255-268},
Volume = {52},
Year = {1984},
doi = {https://doi.org/10.1080/00268978400101201}}

@article{Cuendet2007a,
Author = {M. A. Cuendet and W. F. Van Gunsteren},
Title={On the Calculation of Velocity-Dependent Properties in Molecular Dynamics Simulations Using the Leapfrog Integration Algorithm},
Journal = jcp,
Pages = {184102/1-9},
Volume = {127},
Year = {2007},
doi = {https://doi.org/10.1063/1.2779878}}

@book{Sloan2008a,
Address = {New York},
Author = {E. D. Sloan and C. Koh},
Edition = {3},
Publisher = {CRC PRess},
Title = {{C}lathrate {H}ydrates of {N}atural {G}ases},
Doi = {https://doi.org/10.1201/9781420008494},
Year = {2008}}

@article{Makino2005a,
Author = {T. Makino and T. Sugahara and K. Ohgaki},
Title={Stability Boundaries of Tetrahydrofuran + Water System},
Journal = jced,
Doi = {10.1021/je0502694},
Pages = {2058-2060},
Volume = {50},
Year = {2005}}

@article{Manakov2003a,
Author = {A. Yu. Manakov and S. V. Goryainov and A. V. Kurnosov and
A. Yu. Likhacheva and Y. A. Dyadin and E. G. Larionov},
Title={Clathrate Nature of the High-Pressure Tetrahydrofuran Hydrate Phase
and Some New Data on the Phase Diagram of the Tetrahydrofuran-Water System at Pressures
up to 3 GPa},
Doi = {10.1021/jp0259925},
Journal = jpcb,
Pages = {7861-7866},
Volume = {107},
Year = {2003}}

@article{Potoff2001a,
Author = {J. J. Potoff and J. I. Siepmann},
Title={Vapor-Liquid Equilibria of Mixtures Containing Alkanes, Carbon Dioxide, and Nitrogen},
Journal = aiche,
Pages = {1676-1682},
Volume = {47},
doi = {https://doi.org/10.1002/aic.690470719},
Year = {2001}}

@incollection{Lemmon2019a,
Author = {Lemmon, E. W. and McLinden, M. O. and Friend, D. G.},
Booktitle = {NIST Chemistry WebBook, NIST Standard
Reference Database Number 69},
Editor = {Linstrom, P. J. and Mallard, W. G.},
Publisher = {National Institute of Standards and Technology: Gaithersburg MD},
Url = {http://webbook.nist.gov},
Title = {Thermophysical
Properties of Fluid Systems},
urldate ={10.05.2021},
Year = {2021}}

@article{Parrinello1981a,
Author = {M. Parrinello and A. Rahman},
Title={Polymorphic Transitions in Single Crystals: A New Molecular Dynamics Method},
Journal = {J. Appl. Phys.},
Pages = {7182-7190},
Volume = {52},
Year = {1981},
doi = {https://doi.org/10.1063/1.328693}}

@article{Conde2010b,
title={Determining the three-phase coexistence line in methane hydrates using
computer simulations},
Author = {M. M. Conde and C. Vega},
Journal = jcp,
Pages = {064507},
Volume = {133},
doi = {https://doi.org/10.1063/1.3466751},
Year = {2010}}

@article{Conde2013a,
title={Note: A simple correlation to locate the three phase coexistence line
in methane-hydrate simulations},
Author = {M. M. Conde and C. Vega},
Journal = jcp,
Pages = {056101},
Volume = {138},
doi = {https://doi.org/10.1063/1.4790647},
Year = {2013}}

@article{Espinosa2014a,
Author = {J. R. Espinosa and C. Vega and E. Sanz},
Title={The Mold Integration Method for the Calculation of the Crystal-Fluid
Interfacial Free Energy from Simulations},
Journal = jcp, 
Pages = {134709-1--134709-15},
Volume = {141},
doi = {https://doi.org/10.1063/1.4896621},
Year = {2014}}

@article{Espinosa2015a,
Author = {J. R. Espinosa and C. Vega and C. Valeriani and E. Sanz},
Title={The Crystal-Fluid Interfacial Free Energy and Nucleation Rate of NaCl
from Different Simulation Methods},
Journal = jcp, 
Pages = {194709-1--194709-12},
Volume = {142},
doi = {https://doi.org/10.1063/1.4921185},
Year = {2015}}

@article{Espinosa2016a,
Author = {J. R. Espinosa and C. Vega and E. Sanz},
Title={Ice - Water Interfacial Free Energy for the TIP4P, TIP4P/2005, TIP4P/Ice,
and mW Models As Obtained from the Mold Integration Technique},
Journal = jpcc, 
Pages = {8068-8075},
Volume = {120},
doi = {https://doi.org/10.1021/acs.jpcc.5b11221},
Year = {2016}}

@article{Lechner2008a,
Author = {W. Lechner and C. Dellago},
Title={Accurate Determination of Crystal Structures Based on Averaged Local Bond
Order Parameters},
Journal = jcp, 
Pages = {114707-1--114707-6},
Volume = {129},
doi = {https://doi.org/10.1063/1.2977970},
Year = {2008}}

@article{Uchida1999a,
Author = {T. Uchida and T. Ebinuma and T. Ishizaki},
Title={Dissociation Condition Measurements of Methane Hydrate in Confined Small Pores of Porous Glass},
Journal = jpcb,
Pages = {3659-3662},
Volume = {103},
Doi = {https://doi.org/10.1021/jp984559l},
Year = {1999}}

@article{Uchida2002a,
Author = {T. Uchida and T. Ebinuma and S. Takeya and J. Nagao and H. Narita},
Title={Effects of Pore Sizes on Dissociation Temperatures and Pressures of Methane,
Carbon Dioxide, and Propane Hydrates in Porous Media},
Journal = jpcb,
Pages = {802-826},
Volume = {106},
Doi = {https://doi.org/10.1021/jp012823w},
Year = {2002}}

@article{Anderson2003a,
Author = {R. Anderson and M. Llamedo and B. Tohihi and R. W. Burgass},
Title={Characteristics of Clathrate Hydrate Equilibria in Mesopores and
Interpretation of Experimental Data},
Journal = jpcb,
Pages = {3500-3506},
Volume = {107},
Doi = {https://doi.org/10.1021/jp0263368},
Year = {2003}}

@article{Anderson2003b,
Author = {R. Anderson and M. Llamedo and B. Tohihi and R. W. Burgass},
Title={Experimental Measurement of Methane and Carbon Dioxide Clathrate Hydrate Equilibria
in Mesoporous Silica},
Journal = jpcb,
Pages = {3507-3514},
Volume = {107},
Doi = {https://doi.org/10.1021/jp0263370},
Year = {2003}}

@book{Adamson1997a,
Author = {A. W. Adamson and A. P. Gast},
Publisher = {John Wiley and Sons, Inc.},
Title = {Physical Chemistry of Surfaces},
Year = {1997}}

@book{Debenedetti1996a,
Author = {P. G. Debenedetti},
Publisher = {Princeton University Press},
Title = {Metastable Liquids: Concepts and Principles},
Year = {1997}}

@article{Kvenvolden1988a,
Author = {K. A. Kvenvolden},
Title={Methane Hydrate -- A Major Reservoir of Carbon in the Shallow Geosphere?},
Journal = cg,
Pages = {41-51},
Volume = {71},
Doi = {https://doi.org/10.1016/0009-2541(88)90104-0},
Year = {1988}}

@article{Koh2012a,
Author = {C. A. Koh and A. K. Sum and E. D. Sloan},
Title={State of the Art: Natural Gas Hydrates as a Natural Resource},
Journal = jngse,
Pages = {132-138},
Volume = {8},
Doi = {https://doi.org/10.1016/j.jngse.2012.01.005},
Year = {2012}}

@article{Yang2014a,
Author = {M. Yand and Y. Song and L. Jiang and Y. Zhao and X. Ruan and Y. Zhang and S. Wang},
Title={Hydrate-Based Technology for \ch{CO2} Capture from Fossil Fuel Power Plants},
Journal = ae,
Pages = {26-40},
Volume = {116},
Doi = {https://doi.org/10.1016/j.apenergy.2013.11.031},
Year = {2014}}

@article{Ricaurte2014a,
Author = {M. Ricaurte and C. Dicharry and X. Renaud and J.-P. Torr\'e},
Title={Combination of Surfactants and Organic Compounds for Boosting
\ch{CO2} Separation from Natural Gas by Clathrate Hydrate Formation},
Journal = fuel,
Pages = {206-217},
Volume = {122},
Doi = {https://doi.org/10.1016/j.fuel.2014.01.025},
Year = {2014}}

@article{Kvamme2007a,
Author = {B. Kvamme and A, Graue and T. Buanes and T. Kuznetsova and G. Ersland},
Title={Storage of \ch{CO2} in Natural Gas Hydrate Reservoirs and the Effect of
Hydrate as an Extra Sealing in Cold Aquifers},
Journal = ijggc,
Pages = {236-246},
Volume = {1},
Doi = {https://doi.org/10.1016/S1750-5836(06)00002-8},
Year = {2007}}

@article{Sloan2003a,
Author = {E. D. Sloan},
Title={Fundamental Principles and Applications of Natural Gas Hydrates},
Journal = science,
Pages = {353-359},
Volume = {426},
Doi = {https://doi.org/10.1038/nature02135},
Year = {2003}}

@article{Ohgaki1996a,
Author = {S. Choi and J. H. Drese and C. W. Jones},
Title={Methane Exploitation by Carbon Dioxide from Gas Hydrates—Phase Equilibria
for \ch{CO2}-\ch{CH4} Mixed Hydrate System},
Journal = jcej,
Pages = {478-483},
Volume = {29},
Doi = {https://doi.org/10.1252/jcej.29.478},
Year = {1996}}

@article{Handa1992a,
Author = {Y. P. Handa and D. Stupin},
Title={Thermodynamic Properties and Dissociation Characteristics of Methane and Propane
Hydrates in 70-\AA-Radius Silica Gel Pores},
Journal = jpc,
Pages = {8599-8603},
Volume = {96},
Year = {1992}}

@article{Clennell1999a,
Author = {M. B. Clennell and M. Hovland and J. S. Booth and P. Henry and W. J. Winters},
Title={Formation of Natural Gas Hydrates in Marine Sediments 1. Conceptual Model of
Gas Hydrate Growth Conditioned by Host Sediment Properties},
Journal = jgr,
Pages = {22,985-23,003},
Volume = {104},
doi = {https://doi.org/10.1029/1999JB900175},
Year = {1999}}

@article{Henry1999a,
  title={Formation of natural gas hydrates in marine sediments: 2. Thermodynamic calculations of stability conditions in porous sediments},
  author={Henry, Pierre and Thomas, Michel and Clennell, M Ben},
  journal={Journal of Geophysical Research: Solid Earth},
  volume={104},
  number={B10},
  pages={23005--23022},
  Doi = {https://doi.org/10.1029/1999JB900167},
  year={1999},
  publisher={Wiley Online Library}
}

@article{Zarifi2016a,
Author = {M. Zarifi and J. Javanmardi and H. Hashemi and A. Eslamimanesh and A. H. Mohammadi},
Title={Experimental Study and Thermodynamic Modelling of Methane and Mixed \ch{C1} + \ch{C2} + \ch{C3}
Clathrate Hydrates in the Presence of Mesoporous Silica Gel},
Journal = fpe,
Pages = {17-24},
Volume = {423},
Doi = {https://doi.org/10.1016/j.fluid.2016.03.018},
Year = {2016}}

@article{Kowalsky2007a,
Author = {M. B. Kowalsky and G. J. Moridi},
Title={Comparison of Kinetic and Equilibrium Reaction Models Insimulating Gas Hydrate Behavior in Porous Media},
Journal = ecm,
Pages = {1850-1863},
Volume = {48},
doi = {https://doi.org/10.1016/j.enconman.2007.01.017},
Year = {2007}}

@article{Tsimpanogiannis2019a,
Author = {I. N. Tsimpanigiannis and V. K. Michalis and I. G. Economou},
Title={Enthalpy of Dissociation of Methane Hydrates at a Wide Pressure and Temperature Range},
Journal = fpe,
Pages = {30-40},
Volume = {489},
doi = {https://doi.org/10.1016/j.fluid.2019.01.024},
Year = {2019}}

@article{Churaev1993a,
Author = {N. V. Churaev and S. A. Bardasov and V. D. Sobolev},
Title={On the Non-Freezing Water Interlayers Between Ice and a Silica Surface},
Journal = csa,
Pages = {11-24},
Volume = {79},
doi = {https://doi.org/10.1016/0927-7757(93)80155-8},
Year = {1993}}

@article{Seo2002a,
Author = {Y. Seo and H. Lee and T. Uchida},
Title={Methane and Carbon Dioxide Hydrate Phase Behavior in Small Porous Silicia Gels: Three-Phase
Equilibrium Determination and Thermodynamic Modeling},
Journal = la,
Pages = {9164-9170},
Volume = {18},
Doi = {https://doi.org/10.1021/la0257844},
Year = {2002}}

@article{Kang2008a,
Author = {S.-P. Kang and J.-W- Lee and H.-J. Ryu},
Title={Phase Behavior of Methane and Carbon Dioxide Hydrates in Meso- and Macro-Sized Porous Media},
Journal = fpe,
Pages = {68-72},
Volume = {274},
Doi= {https://doi.org/10.1016/j.fluid.2008.09.003},
Year = {2008}}

@article{Smith2002a,
Author = {D. H. Smith and J. W. Wilder and K. Seshadri},
Title={Methane Hydrate Equilibria in Silica Gels with Broad Pore-Size Distributions},
Journal = aiche,
Pages = {393-400},
Volume = {48},
Doi = {https://doi.org/10.1002/aic.690480222},
Year = {2002}}

@article{Smith2004a,
Author = {D. H. Smith and K. Seshadri and T. Uchida and J. W. Wilder},
Title={Thermodynamics of Methane, Propane, and Carbon Dioxide Hydrates in Porous Glass},
Journal = aiche,
Pages = {1589-1598},
Volume = {50},
Doi = {https://doi.org/10.1002/aic.10141},
Year = {2004}}

@article{Zhang2003a,
Author = {W. Zhang and J. W. Wilder and D. H. Smith},
Title={Methane Hydrate-Ice Equilibria in Porous Media},
Journal = jpcb,
Pages = {13084-13089},
Volume = {107},
Doi = {https://doi.org/10.1021/jp0349278},
Year = {2003}}

@article{Zhang2016a,
Author = {Y. Zhang and X.-S. Li and Y. Wang and Z.-Y. Chen and K.-F- Yan},
Title={Decomposition Conditions of Methane Hydrate in Marine Sediments from South China Sea},
Journal = fpe,
Pages = {110-115},
Volume = {413},
Doi = {https://doi.org/10.1016/j.fluid.2015.12.004},
Year = {2016}}

@article{Liu2018a,
Author = {H. Liu and S. Zhan and P. Guo and S. Fan and S. Zhang},
Title={Understanding the Characteristic of Methane Hydrate Equilibrium in T Materials and Its Potential Application},
Journal = cej,
Pages = {775-781},
Volume = {349},
Doi = {https://doi.org/10.1016/j.cej.2018.05.150},
Year = {2018}}

@article{Ishikiriyama1995a,
Author = {K. Ishikiriyama and M. Todoki and K. Motomura},
Title={Pore Size Distribution (PSD) Measurements of Silica Gel by Means of Differential Scanning
Calorimetry},
Journal = jcis,
Pages = {92-102},
Volume = {171},
doi = {https://doi.org/10.1006/jcis.1995.1155},
Year = {1995}}

@article{Faivre1999a,
Author = {C. Faivre and D. Bellet and G. Dolino},
Title={Phase Transitions of Fluids Confined in Porous Silicon: A Differential Calorimetry Investigation},
Journal = epjb,
Pages = {19-36},
Volume = {7},
doi = {https://doi.org/10.1007/s100510050586},
Year = {1999}}

@article{Radhakrishnan2002a,
Author = {R. Radhakrishnan and K. E. Gubbins and M. Sliwinska-Bartkowiak},
Title={Global Phase Diagrams for Freezing in Porous Media},
Journal = jcp,
Pages = {1147-1155},
Volume = {116},
doi = {https://doi.org/10.1063/1.1426412},
Year = {2002}}

@article{Broughton1986a,
Author = {J. Q. Broughton and G. H. Gilmer},
Title={Molecular Dynamics Investigation of the Crystal-Fluid Interface. VI. Excess
Surface Free Energies of Crystal-Liquid Systems},
Journal = jcp,
Pages = {5759-5768},
Volume = {84},
Doi = {https://doi.org/10.1063/1.449884},
Year = {1986}}

@article{Fernandez2012a,
Author = {L. A. Fernandez and V. Martin-Mayor and B. Seoane and P. Verrocchio},
Title={Equilibrium Fluid-Solid Coexistence of Hard Spheres},
Journal = prevlet,
Pages = {165701-1--165701-5},
Volume = {108},
Year = {2012}}

@article{Angioletti-Uberti2010a,
Author = {S. Angioletti-Uberti and M. Ceriotti and P. D. Lee and M. W. Finnis},
Title={Solid-Liquid Interface Free Energy Through Metadynamics Simulations},
Journal = prb,
Pages = {125416},
Volume = {81},
Doi = {https://doi.org/10.1103/PhysRevB.81.125416},
Year = {2010}}

@article{Hoyt2001a,
Author = {J. J. Hoyt and M. Asta and A. Karma},
Title={Method for Computing the Anisotropy of the Solid-Liquid Interfacial Free Energy},
Journal = prevlet,
Pages = {5530-5533},
Volume = {86},
Doi = {https://doi.org/10.1103/PhysRevLett.86.5530},
Year = {2001}}

@article{Knott2012a,
Author = {B. C. Knott and V. Molinero and M. F. Doherty and B. Peters},
Title={Homogeneous Nucleation of Methane Hydrates: Unrealistic Under Realistic Conditions},
Journal = jacs,
Pages = {19544-19547},
Volume = {134},
doi = {https://doi.org/10.1021/ja309117d},
Year = {2012}}

@article{Jacobson2010a,
Author = {L. C. Jacobson and W. Hujo and V. Molinero},
Title={Amorphous Precursors in the Nucleation of Clathrate Hydrates},
Journal = jacs,
Pages = {11806-11811},
Volume = {132},
Doi = {https://doi.org/10.1021/ja1051445},
Year = {2010}}

@article{Jacobson2010b,
Author = {L. C. Jacobson and W. Hujo and V. Molinero},
Title={Nucleation Pathways of Clathrate Hydrates: Effect of Guest Size and Solubility},
Journal = jpcb,
Pages = {13796-13807},
Volume = {114},
Doi = {https://doi.org/10.1021/jp107269q},
Year = {2010}}

@article{Jacobson2011a,
Author = {L. C. Jacobson and V. Molinero},
Title={Can Amorphous Nuclei Grow Crystalline Clathrates? The Size and Crystallinity of Critical Clathrate Nuclei},
Journal = jacs,
Pages = {6458-6463},
Volume = {133},
doi = {https://doi.org/10.1021/ja201403q},
Year = {2011}}

@article{Sanz2013a,
Author = {E. Sanz and J. R. Espinosa and R. Caballero-Bernal and J. L. F. Abascal and C. Valeriani},
Title={Homogeneous Ice Nucleation at Moderate Supercooling from Molecular Simulation},
Journal = jacs,
Pages = {15008-15017},
Volume = {135},
doi ={https://doi.org/10.1021/ja4028814},
Year = {2013}}

@article{Constandy2015a,
Author = {J. Costandy and V. K. Michalis and I. N. Tsimpanogiannis and A. K. Stubos and I. G. Economou},
Title={The Role of Intermolecular Interactions in the Prediction of the Phase Equilibria of Carbon Dioxide Hydrates},
Journal = jcp,
Pages = {094506-1--094506-8},
Volume = {143},
doi = {https://doi.org/10.1063/1.4929805},
Year = {2015}}

@article{Waage2017a,
Author = {M. H. Waage and T. J. H. Vlugt and S. Kjelstrup},
Title={Phase Diagram of Methane and Carbon Dioxide Hydrates Computed by Monte Carlo Simulations},
Journal = jpcb,
Pages = {7336-7350},
Volume = {121},
doi = {https://doi.org/10.1021/acs.jpcb.7b03071},
Year = {2017}}

@article{Bai2006a,
Author = {X.-M. Bai and M. Li},
Title={Calculation of Solid-Liquid Interfacial Free Energy: A Classical Nucleation Theory Based Approach},
Journal = jcp,
Pages = {124707-1--124707-12},
Volume = {124},
doi = {https://doi.org/10.1063/1.2184315},
Year = {2006}}

@article{Mirzaeifard2019a,
Author = {S. Mirzaeifard and P. Servio and A. D. Rey},
Title={Characterization of Nucleation of Methane Hydrate Crystals: Interfacial Theory and Molecular Simulation},
Journal = jcis,
Pages = {556-567},
Volume = {557},
doi = {https://doi.org/10.1016/j.jcis.2019.09.056},
Year = {2019}}

@article{Mirzaeifard2019b,
Author = {S. Mirzaeifard and P. Servio and A. D. Rey},
Title={Multiscale Modeling and Simulation of Water and Methane Hydrate Crystal Interface},
Journal = cgd,
Pages = {5142-5151},
Volume = {19},
doi = {https://doi.org/10.1021/acs.cgd.9b00578},
Year = {2019}}

@article{Naeiji2017a,
Author = {P. Naeiji and F. Varaminian adn M. Rahmati},
Title={Comparison of the Thermodynamic, Structural and Dynamical Properties of
Methane/Water and Methane/Water/Hydrate Systems Using Molecular Dynamic Simulations},
Journal = jngse,
Pages = {122-130},
Volume = {44},
doi = {https://doi.org/10.1016/j.jngse.2017.04.010},
Year = {2017}}

@article{Abascal2005b,
Author = {J. L. F. Abascal and E. Sanz and R. Garc{\'{\i}}a Fern\'andez and C. Vega},
Title={A Potential Model for the Study of Ices and Amorphous Water: TIP4P/Ice},
Journal = jcp,
Pages = {234511-1--234511-9},
Volume = {122},
doi = {https://doi.org/10.1063/1.1931662},
Year = {2005}}

@article{Reinhardt2012a,
Author = {A. Reinhardt and J. P. Doye and E. G. Noya and C. Vega},
Title={Local Order Parameters for Use in Driving Homogeneous Ice Nucleation with All-Atom Models of Water},
Journal = jcp,
Pages = {194504-1--194594-9},
Volume = {137},
doi = {https://doi.org/10.1063/1.4766362},
Year = {2012}}

@article{Nakano1998a,
Author = {S. Nakano and M. Moritoki and K. Ohgaki},
Title={High-Pressure Phase Equilibrium and Raman Microprobe
Spectroscopic Studies on the \ch{CO2} Hydrate System},
Journal = jced,
Pages = {807-810},
Volume = {43},
Year = {1998},
doi = {https://doi.org/10.1021/je9800555}}

@article{Schreiber2001a,
Author = {A. Schreiber and I. Ketelsen and G. H. Findenegg},
Title={Melting and Freezing of Water in Ordered Mesoporous Silica Materials},
Journal = pccp,
Pages = {1185-1195},
Volume = {3},
doi = {https://doi.org/10.1039/B010086M},
Year = {2001}}

@article{Jahnert2008a,
Author = {S. J\"ahnert and F. Vaca Ch\'avez and G. E. Schaumann and A. Schreiber and
M. Sch\"onhoff and G. H. Findenegg},
Title={Melting and Freezing of Water in Cylindrical Silica Nanopores},
Journal = pccp,
Pages = {6039-6051},
Volume = {10},
doi = {https://doi.org/10.1039/B809438C},
Year = {2008}}

@article{Yao2019a,
Author = {Y. Yao and V. Fella and W. Huang and K. A. I. Zhang and K. Landfester and
H.-J. Butt and M. Vogel and G. Floudas},
Title={Crystallization and Dynamics of Water Confined in Model Mesoporous Silica
Particles: Two Ice Nuclei and Two Fractions of Water},
Journal = la,
Pages = {5890-5901},
Volume = {35},
doi = {https://doi.org/10.1021/acs.langmuir.9b00496},
Year = {2019}}

@article{Kittaka2011a,
Author = {S. Kittaka and Y. Ueda and F. Fujisaki and T. Iiyamab and T. Yamaguchi},
Title={Mechanism of Freezing of Water in Contact with Mesoporous Silicas MCM-41, SBA-15 and SBA-16:
Role of Boundary Water of Pore Outlets in Freezing},
Journal = pccp,
Pages = {17222-17233},
doi ={https://doi.org/10.1039/C1CP21458F},
Volume = {13},
Year = {2011}}

@article{Morishige2018a,
Author = {K. Morishige},
Title={Influence of Pore Wall Hydrophobicity on Freezing and Melting of Confined Water},
Journal = jpcc,
Pages = {5013-5019},
Volume = {122},
doi = {https://doi.org/10.1021/acs.jpcc.8b00538},
Year = {2018}}

@article{Davidchack2000a,
Author = {R. L. Davidchack and B. B. Laird},
Title={Direct Calculation of the Hard-Sphere Crystal/Melt Interfacial Free Energy},
Journal = prevlet,
Pages = {4752-4754},
Volume = {85},
Doi = {https://doi.org/10.1103/PhysRevLett.85.4751},
Year = {2000}}

@article{Davidchack2003a,
Author = {R. L. Davidchack and B. B. Laird},
Title={Direct Calculation of the Crystal–Melt Interfacial Free Energies
for Continuous Potentials: Application to the Lennard-Jones System},
Journal = jcp,
Pages = {7651-7657},
Volume = {118},
doi = {https://doi.org/10.1063/1.1563248},
Year = {2003}}

@article{Handel2008a,
Author = {R. Handel and  R. L. Davidchack and J. Anwar and A. Brukhno},
Title={Direct Calculation of Solid-Liquid Interfacial Free Energy for Molecular Systems:
 TIP4P Ice-Water Interface},
Journal = prevlet,
Pages = {036104-1--036104-4},
Volume = {100},
doi = {https://doi.org/10.1103/PhysRevLett.100.036104},
Year = {2008}}

@article{Davidchack2010a,
Author = {R. L. Davidchack},
Title={Hard Spheres Revisited: Accurate Calculation of the Solid–Liquid Interfacial Free Energy},
Journal = jcp,
Pages = {234701-1--234701-7},
Volume = {133},
Doi = {https://doi.org/10.1063/1.3514144},
Year = {2010}}

@article{Davidchack2012a,
Author = {R. L. Davidchack and R. Handel and J. Anwar and A. V. Brukhno},
Title={Ice Ih-Water Interfacial Free Energy of Simple Water Models with Full Electrostatic Interactions},
Journal = jctc,
Pages = {2383-2390},
Volume = {8},
doi = {https://doi.org/10.1021/ct300193e},
Year = {2012}}

@article{Chihaia2005a,
Author = {V. Chihaua and S. Adams and W. F. Kuhs},
Title={Molecular dynamics simulations of properties of a (001) methane clathrate hydrate surface},
Journal = "Chem. Phys.",
Pages = {208-225},
Volume = {317},
Doi = {https://doi.org/10.1016/j.chemphys.2005.05.024},
Year = {2005}}

@article{Peters2008a,
Author = {B. Peters and N. E. R. Zimmermann and G. T. Beckham and J. W. Tester and B. J. Trout},
Title={Path Sampling Calculation of Methane Diffusivity in
Natural Gas Hydrates from a Water-Vacancy Assisted Mechanism},
Journal = jacs,
Pages = {17342-17350},
Volume = {130},
Doi= {https://doi.org/10.1021/ja802014m},
Year = {2008}}

@article{English2009a,
Author = {N. J. English and J. S. Tse},
Title={Mechanisms for Thermal Conduction in Methane Hydrate},
Journal = prevlet,
Pages = {015901-1--015901-4},
Volume = {103},
Doi = {https://doi.org/10.1103/PhysRevLett.103.015901},
Year = {2009}}

@article{vanDuijneveldt1992a,
Author = {J. S. van Duijneveldt and D. Frenkel},
Title={Computer simulation study of free energy barriers in crystal nucleation},
Journal = jcp,
Pages = {4655--4668},
Volume = {96},
doi = {https://doi.org/10.1063/1.462802},
Year = {1992}}

@article{Lynden-Bell1993a,
Author = {R. M. Lynden-Bell and J. S. van Duijneveldt and D. Frenkel},
Title={Free energy changes on freezing and melting ductile metals},
Journal = molphys,
Pages = {801--814},
Volume = {80},
doi = {https://doi.org/10.1080/00268979300102661},
Year = {1993}}

@article{tenWolde1995a,
Author = {P. R. ten Wolde and M. J. Ruiz-Montero and D. Frenkel},
Title={Numerical Evidence for bcc Ordering at the Surface of a Critical fcc Nucleus},
Journal = prevlet,
Pages = {2714--2717},
Volume = {75},
doi = {https://doi.org/10.1103/PhysRevLett.75.2714},
Year = {1995}}

@article{tenWolde1996a,
Author = {P. R. ten Wolde and M. J. Ruiz-Montero and D. Frenkel},
Title={Numerical calculation of the rate of crystal nucleation in a
Lennard-Jones system at moderate undercooling},
Journal = jcp,
Pages = {9932--9947},
Volume = {104},
doi = {https://doi.org/10.1063/1.471721},
Year = {1996}}

@article{tenWolde1997a,
Author = {P. R. ten Wolde and D. Frenkel},
Title={Enhancement of Protein Crystal Nucleation by Critical Density Fluctuations},
Journal = science,
Pages = {1975--1978},
Volume = {277},
doi = {http://dx.doi.org/10.1126/science.277.5334.1975},
Year = {1997}}

@article{Aman2016a,
  title={Interfacial phenomena in gas hydrate systems},
  author={Aman, Zachary M and Koh, Carolyn A},
  journal={Chemical Society Reviews},
  volume={45},
  number={6},
  pages={1678--1690},
  Doi = {10.1039/C5CS00791G},
  year={2016},
  publisher={Royal Society of Chemistry}
}

@incollection{evans200515,
  title={15 Measurement of surface and interfacial tension},
  author={Evans, MJB},
  booktitle={Experimental Thermodynamics},
  volume={7},
  pages={383--407},
  doi = {https://doi.org/10.1016/S1874-5644(05)80017-7},
  year={2005},
  publisher={Elsevier}
}

@article{truskett2000,
  title={Towards a quantification of disorder in materials: Distinguishing equilibrium and glassy sphere packings},
  author={Truskett, TM and Torquato, S and Debenedetti, PG},
  journal={Physical Review E},
  volume={62},
  number={1},
  pages={993},
  year={2000},
  doi = {https://doi.org/10.1103/PhysRevE.62.993},
  publisher={APS}
}

@article{chau1998,
  title={A new order parameter for tetrahedral configurations},
  author={Chau, P-L and Hardwick, AJ},
  journal={Molecular Physics},
  volume={93},
  number={3},
  pages={511--518},
  year={1998},
  doi = {https://doi.org/10.1080/002689798169195},
  publisher={Taylor \& Francis}
}

@article{sebastiani2020,
  title={An isolated water droplet in the aqueous solution of a supramolecular tetrahedral cage},
  author={Sebastiani, Federico and Bender, Trandon A and Pezzotti, Simone and Li, Wan-Lu and Schwaab, Gerhard and Bergman, Robert G and Raymond, Kenneth N and Toste, F Dean and Head-Gordon, Teresa and Havenith, Martina},
  journal={Proceedings of the National Academy of Sciences},
  volume={117},
  number={52},
  pages={32954--32961},
  year={2020},
  doi = {https://doi.org/10.1073/pnas.2012545117},
  publisher={National Acad Sciences}
}

@article{Liu2011a,
title={Monte Carlo Simulations of High-Pressure Phase Equilibria of \ch{CO2}--\ch{H2O} Mixtures},
author={Y. Liu and A. Z. Panagiotopoulos and P. G. Debenedetti},
journal=jpcb,
volume={115},
pages={6629-6635},
doi = {https://doi.org/10.1021/jp201520u},
year={2011}
}

@article{Orozco2014a,
title={Optimization of Intermolecular Potential Parameters for the \ch{CO2}/\ch{H2O} Mixture},
author={G. A. Orozco and I. G. Economou and A. Z. Panagiotopoulos},
journal=jpcb,
volume={118},
pages={11504-11511},
doi = {https://doi.org/10.1021/jp5067023},
year={2014}
}

@article{Yu2012a,
title={Many-body effects are essential in a physically motivated \ch{CO2} force field},
author={K. Yu and J. R. Schmidt},
journal=jcp,
volume={136},
pages={034503-1–-034503-7},
doi = {https://doi.org/10.1063/1.3672810},
year={2012}
}

@article{Riera2020a,
title={Data-Driven Many-Body Models for Molecular Fluids: \ch{CO2}/\ch{H2O} Mixtures as a Case Study},
author={M. Riera and E. P. Yeh and F. Paesani},
journal=jctc,
volume={16},
pages={2246-2257},
doi = {https://doi.org/10.1021/acs.jctc.9b01175},
year={2020}
}

@article{Murthy1981a,
title={Interaction site models for carbon dioxide},
Author = {C. S. Murthy and K. Singer and I. R. McDonald},
Journal = {Mol. Phys.},
Pages = {13},
Volume = {44},
doi = {https://doi.org/10.1080/00268978100102331},
Year = {1981}}

@article{Moller1994a,
title={Determination of an effective intermolecular potential for
carbon dioxide using vapour-liquid phase equilibria from
NpT + test particle simulations},
Author = {D. Moller and J. Fischer},
Journal = {Fluid Phase Equilib.},
Pages = {35},
Volume = {100},
doi = {https://doi.org/10.1016/0378-3812(94)80002-2},
Year = {1994}}

@article{Murthy1983a,
title={Electrostatic interactions in molecular crystals},
Author = {C. S. Murthy and S. F. Oshea and I. R. Mc Donald},
Journal = {Mol. Phys.},
Pages = {531},
Volume = {50},
doi = {https://doi.org/10.1080/00268978300102531},
Year = {1983}}

@article{Harris1995a,
title={Carbon Dioxide's Liquid-Vapor Coexistence Curve And Critical Properties as Predicted by a Simple Molecular Model},
Author = {J. G. Harris and K. H. Yung},
Journal = {J. Phys. Chem.},
Pages = {12021},
Volume = {99},
doi = {https://doi.org/10.1021/j100031a034},
Year = {1995}}

@article{Zhang2005a,
title={An optimized molecular potential for carbon dioxide},
Author = {Z. Zhang and Z. Duan},
Journal = {J. Chem. Phys.},
Pages = {214507},
Volume = {122},
doi = {https://doi.org/10.1063/1.1924700},
Year = {2005}}

@article{Perez-Sanchez2013a,
title={Fluid-solid equilibrium of carbon dioxide as obtained from computer simulations of several popular potential models: The role of the quadrupole},
Author = {G. Per{\'e}z-Sanchez and D. Gonz{\'a}lez-Salgado and M. M. Pi{\~n}eiro and C. Vega},
Journal = {J. Chem. Phys.},
Pages = {084506},
Volume = {138},
doi = {http://dx.doi.org/10.1063/1.4792443},
Year = {2013}}

@article{Fernandez-Fernandez2019a,
title={Three-phase equilibrium curve shift for methane hydrate in oceanic conditions calculated from Molecular Dynamics simulations},
Author = {A. M. Fern{\'a}ndez-Fern{\'a}ndez and M. P{\'e}rez-Rodr{\'{\i}}guez and A. Comesa{\~n}a and M. M. Pi{\~n}eiro},
Journal = {J. Mol. Liquids},
Pages = {426},
Volume = {274},
doi = {https://doi.org/10.1016/j.molliq.2018.10.146},
Year = {2019}}

@article{Fernandez-Fernandez2021a,
title={Molecular dynamics of fluoromethane type I hydrates},
Author = {A. M. Fern{\'a}ndez-Fern{\'a}ndez and M. P{\'e}rez-Rodr{\'{\i}}guez and M. M. Pi{\~n}eiro},
Journal = {J. Mol. Liquids},
Pages = {116720},
Volume = {339},
doi = {https://doi.org/10.1016/j.molliq.2021.116720},
Year = {2021}}

@article{Benet2014a,
title={Computer simulation study of surface wave dynamics at the crystal-melt interface},
Author = {J. Benet and L. G. MacDowell and E. Sanz},
Journal = jcp,
Pages = {034701},
Volume = {141},
doi = {https://doi.org/10.1063/1.4886806},
Year = {2014}}

@article{Phan2022a,
title={Surface morphology effects on clathrate hydrate wettability},
Author = {A. Phan and H. M. Stoner and M. Stamatakis and C. A. Koh and A. Striolo},
Journal = jcis,
Pages = {421-431},
Volume = {611},
doi = {https://doi.org/10.1016/j.jcis.2021.12.083},
Year = {2022}}

@article{Phan2021a,
title={Molecular mechanisms by which tetrahydrofuran affects CO$_{2}$ hydrate Growth: Implications for carbon storage},
Author = {A. Phan and H. Schl{\"o}sser and A. Striolo},
Journal = cej,
Pages = {129423},
Volume = {418},
doi = {https://doi.org/10.1016/j.cej.2021.129423},
Year = {2021}}

@article{Majid2021a,
title={Phase behavior and kinetics properties of gas hydrates in confinement and its application},
Author = {A. A. A. Majid and C. A. Koh},
Journal = aiche,
Pages = {e17176},
Volume = {67},
doi = {https://doi.org/10.1002/aic.17176},
Year = {2021}}

@article{Hu2017a,
title={Interfacial Properties and Mechanisms Dominating Gas Hydrate
Cohesion and Adhesion in Liquid and Vapor Hydrocarbon Phases},
Author = {S. Hu and C. A. Koh},
Journal = la,
Pages = {11299-11309},
Volume = {33},
doi = {https://doi.org/10.1021/acs.langmuir.7b02676},
Year = {2017}}

@article{Hu2021a,
title={Structural Effects of Gas Hydrate Antiagglomerant Molecules on Interfacial Interparticle Force Interactions},
Author = {S. Hu and L. Vo and D. Monteiro and S. Bodnar and P. Prince and C. A. Koh},
Journal = la,
Pages = {1651-1661},
Volume = {37},
doi = {https://dx.doi.org/10.1021/acs.langmuir.0c02503},
Year = {2021}}

@article{Reddy2016a,
title={On the accuracy of the MB-pol many-body potential for water: Interaction energies, vibrational frequencies, and classical thermodynamic and dynamical properties from clusters to liquid water and ice},
Author = {S. K. Reddy and S. C. Stright and P. Bajaj and C. H. Pham and M. Riera and D. R. Moberg and M. A. Morales and C. Knight and A. W. G\"otz and F, Paesani},
Journal = jcp,
Pages = {194504},
Volume = {145},
doi = {https://doi.org/10.1063/1.4967719},
Year = {2019}}

%% else use the following coding to input the bibitems directly in the
%% TeX file.

% \begin{thebibliography}{00}

% %% \bibitem{label}
% %% Text of bibliographic item

% \bibitem{}

% \end{thebibliography}
\end{document}